\documentclass[10pt,twocolumn, a4paper, nofootinbib]{article}

\usepackage[english]{babel}   % "babel.sty" + "french.sty"
\usepackage[nocompress]{cite}
\usepackage{wacv_arXiv}
\usepackage{times}
\usepackage{epsfig}
\usepackage{graphicx}
\usepackage{amsmath}
\usepackage{amssymb}
\usepackage{booktabs}
\usepackage{url}
\usepackage{libertine}
\usepackage{xcolor}
\usepackage{subcaption}
\usepackage{multirow}

\usepackage{tikz}
\usetikzlibrary{calc,spy}
\usepackage{pgfplots}
\usepackage{pgfkeys}
\usetikzlibrary{spy}
\usetikzlibrary[patterns]
\usetikzlibrary{calc,decorations.pathreplacing}
\usetikzlibrary{shadows.blur}
\usetikzlibrary{shapes.symbols}
\tikzstyle{closeup} = [
opacity=1.0,          % opacity of both windows
height=0.075 \linewidth,         %  height of large window
width=0.075 \linewidth,          %   width of large window
connect spies, green  % connects big and small window by a green line
]
\tikzstyle{largewindow} = [red, line width=0.50mm]
\tikzstyle{smallwindow} = [blue,line width=0.20mm]

\usepackage{color}
\usepackage[switch]{lineno}
\usepackage[vlined,ruled]{algorithm2e}
\usepackage[T1]{fontenc}

\wacvalgorithmstrack   % Uncomment this line if you are submitting to the Algorithms Track.
\wacvfinalcopy % *** Uncomment this line for the final submission

\ifwacvfinal
\usepackage[breaklinks=true,bookmarks=false]{hyperref}
\else
\usepackage[pagebackref=true,breaklinks=true,colorlinks,bookmarks=false]{hyperref}
\fi

\begin{document}
\sloppy

%%%%%%%%% TITLE
\title{Video Restoration with a Deep Plug-and-Play Prior}

\author{Antoine Monod\\
\small MAP5 / GoPro Inc.\\
{\tt\scriptsize amonod@gopro.com}
\and
Julie Delon\\
\small MAP5\\
{\tt\scriptsize julie.delon@u-paris.fr}
\and
Matias Tassano\\
\small Meta Inc.\thanks{Work mostly done while Matias was at GoPro Inc.}\\
{\tt\scriptsize tasso.matias@gmail.com}
\and
Andrés Almansa\\
\small MAP5\\
{\tt\scriptsize andres.almansa@u-paris.fr}
}
\maketitle
%\thispagestyle{empty}

%%%%%%%%% ABSTRACT
\begin{abstract}
	This paper presents a novel method for restoring digital videos via a Deep Plug-and-Play (PnP) approach.
	Under a Bayesian formalism, the method consists in using a deep convolutional denoising network
	in place of the proximal operator of the prior in an alternating optimization scheme.
	We distinguish ourselves from prior PnP work by directly applying that method
	to restore a digital video from a degraded video observation.
	This way, a network trained once for denoising can be repurposed for other video restoration tasks. 
	Our experiments in video deblurring, super-resolution, and interpolation of random missing pixels 
	all show a clear benefit to using a network specifically designed for video denoising, 
	as it yields better restoration performance and better temporal stability
	than a single image network with similar denoising performance using the same PnP formulation.
	Moreover, our method compares favorably to applying a different state-of-the-art PnP scheme separately on each frame of the sequence.
	This opens new perspectives in the field of video restoration.
\end{abstract}

%%%%%%%%% BODY TEXT
	\section{Introduction}
	
	%\subsection{Formalisme bayésien pour la restauration de séquences d'images}
	
	Video restoration tasks, like their single image counterparts, can in many cases be seen as inverse problems, whose direct form can be written as
	\begin{equation}
		\mathbf{y} = \mathcal{A}(\mathbf{x}) + \mathbf{n}
		\label{eq:forward}
	\end{equation}
	where $\mathbf{y} \in \mathbb{R}^m$ is the degraded observation, $\mathbf{x} \in \mathbb{R}^d$ is the unknown video or image to be recovered, $\mathcal{A}$ is a degradation operator that is often linear, and $\mathbf{n}  \in \mathbb{R}^m$ is a noise realization from a known distribution.
	%$n \sim \mathcal{N}(0,\sigma_{\mathbf{n}}^2)$.
	Although the distribution of noise observed  in real images is typically not uniquely Gaussian~\cite{foi_practical_nodate, wei_physics-based_2020}, such a model is frequently used in the image processing literature, as it features mathematical properties that are key to many methods, including the one presented in this work. These inverse problems are often ill-posed or at least ill-conditioned. In order to build reliable and robust estimators of $\mathbf{x}$ from the observation $\mathbf{y}$, it is common to use a Bayesian formalism, in which one assumes that the unknown $\mathbf{x}$ follows a law of density $p(\mathbf{x})$ (called the "prior"). By combining it with $p(\mathbf{y}|\mathbf{x})$, the likelihood of $\mathbf{y}$ knowing $\mathbf{x}$ (given by the degradation model (1)), we obtain the posterior density $p(\mathbf{x}|\mathbf{y})$, whose maximum a posteriori (MAP) is generally sought:
	\begin{equation}\label{eq:bayes}
		\hat{\mathbf{x}} = \underset{\mathbf{x}}{\operatorname{argmax}}\;p(\mathbf{x}|\mathbf{y}) =  \underset{\mathbf{x}}{\operatorname{argmax}} \log p(\mathbf{y}|\mathbf{x}) + \log p(\mathbf{x}).
	\end{equation}
	% et ne dépend pas de l'observation $\mathbf{y}$".
	In the case where the noise is Gaussian i.i.d. of variance $\sigma_{\mathbf{n}}^{2}$, the problem can be rewritten in variational form
	\begin{equation}\label{eq:energy_minimization}
		\hat{\mathbf{x}} = \underset{\mathbf{x}}{\operatorname{argmin}} \frac{1}{2 \sigma_{\mathbf{n}}^{2}}\|\mathbf{y}-\mathcal{A}(\mathbf{x})\|_{2}^{2} + \alpha\mathcal{R}(\mathbf{x}),
	\end{equation}
	where the log-likelihood of the observation (also called data fidelity term) is $-\frac{1}{2 \sigma_{\mathbf{n}}^{2}}\|\mathbf{y}-\mathcal{A}(\mathbf{x})\|_{2}^{2}$ , and the log-prior on the unknown (also called regularization term) is  $-\alpha\mathcal{R}(\mathbf{x})$.
	
	For a long time, Bayesian restoration in digital imaging or video has relied on explicit priors (such as total variation~\cite{rudin_nonlinear_1992}), expressing regularity assumptions on $\mathbf{x}$ either in the image space or in transformed spaces (wavelet transforms, patch spaces, etc.)~\cite{bertalmio_gaussian_2018}. When $\mathcal{R}$ is known and convex, there are numerous efficient numerical schemes to find the solutions of~\eqref{eq:energy_minimization}~\cite{chambolle2016introduction}.
	
	In recent years, deep neural networks have outperformed these traditional restoration methods for most image and video restoration problems. The so-called \textit{end-to-end} networks (see~\cite{zhang_ffdnet_2018, brooks_unprocessing_2018, tassano_fastdvdnet_2020, zamir_multi-stage_2021} for examples of efficient image or video denoising networks) are directly trained from pairs $(\mathbf{x_i}, \mathbf{y_i})$ satisfying the degradation model~\eqref{eq:forward}. Training these networks requires large amounts of data and computing resources (and these, along with network size, tend to keep increasing over the years). Moreover, a network trained for a given degradation model must be retrained as soon as the degradation model or its parameters change.
	
	\textit{Plug-and-Play} (PnP) methods try to bridge some of the gaps between these two approaches. They combine a likelihood defined explicitly according to the direct model~\eqref{eq:forward}, and a prior $\mathcal{R}$ implicitly defined by an efficient denoising algorithm. This combination is done algorithmically, typically within an alternate optimization scheme where the proximal operator of $\mathcal{R}$~\cite{combettes_proximal_2010} (or sometimes its gradient~\cite{becky_fast_2009}) is replaced by the denoiser. PnP methods allow repurposing a single denoising network, trained once, for many restoration tasks. They also allow meeting memory constraints, \eg in an embedded system (like an action camera or a cell phone) where the weights of a single network can be stored for several use cases.
	
	While these PnP approaches have been thoroughly explored for image restoration problems, there have been surprisingly few attempts to use them for video restoration, even though the potential applications in this field are numerous.  In this paper, we show how a deep video denoising network can be used in a PnP scheme to solve different kinds of inverse problems affecting the whole video sequence.  Even for inverse problems affecting video frames separately, such as super-resolution or deblurring, we show that this PnP video scheme compares favorably, in terms of image quality and temporal stability, to applying state-of-the-art PnP schemes on each frame of the sequence. This opens new perspectives in the field of video restoration. The source code of all of our experiments, along with results stored as video files, are available on GitHub ( \url{https://github.com/amonod/pnp-video}).
	
	\section{Related works}
	
	Generally speaking, PnP methods transform the original restoration process into two sub-problems which are easier to solve. The data and regularization/prior terms of the objective function are decoupled by the use of a splitting algorithm. Then, the data and prior sub-problems are solved alternately. As for the prior sub-problem, PnP methods make use of off-the-shelf denoisers to approximate its solution.
	
	\subsection{Plug-and-Play and Deep Plug-and-Play methods}
	PnP methods can be traced back to the seminal work by Venkatakrishnan \textit{et al}.~\cite{venkatakrishnan_plug-and-play_2013} which employs Alternating Direction Method of Multipliers~\cite{boyd_distributed_2011} (or ADMM) optimization to decouple the data and regularization terms. Multiple PnP formulations of optimization algorithms have been proposed, such as ADMM~\cite{venkatakrishnan_plug-and-play_2013, ryu_plug-and-play_2019}, stochastic gradient descent~\cite{laumont_maximum--posteriori_2021}, primal-dual methods~\cite{meinhardt_learning_2017, heide_flexisp_2014}, iterative shrinkage thresholding (ISTA)~\cite{gavaskar_plug-and-play_2020, xu_provable_2020}, fast ISTA~\cite{dpnp_fista} or half quadratic splitting~\cite{zhang_plug-and-play_2021}.
	
	A large diversity of denoisers have been used for the regularization. Among them, BM3D has been used the most~\cite{heide_flexisp_2014, dpnp_bm3d, dpnp_fista}.	
	Most recent PnP methods generally use deep neural networks~\cite{meinhardt_learning_2017, ryu_plug-and-play_2019, laumont_maximum--posteriori_2021}. In~\cite{zhang_plug-and-play_2021}, an analysis of the efficiency of the different deep denoisers for different image restoration tasks is provided. Deep plug-and-play methods can be used to solve different kinds of image resoration tasks such as super-resolution~\cite{brifman_turning_2016}, Gaussian denoising~\cite{dpnp_gauss}, or image deblurring~\cite{dpnp_deblur}. Theoretical aspects of deep plug-and-play algorithms have also been studied using bounded denoisers assumptions~\cite{chan_plug-and-play_2016} or more recently using denoisers whose residual operators are Lipschitz-continuous~\cite{ryu_plug-and-play_2019, laumont_maximum--posteriori_2021}.

	The use of Plug-and-Play operators framework has also been shown to be very efficient with Approximate Message	Passing (AMP) algorithms~\cite{donoho_message-passing_2009,ahmad2020}, particularly for applications involving randomised forward operators, where it is possible
	to characterise AMP schemes in detail (see, e.g.,~\cite{bayati2011,javanmard2013}). The restriction on the forward operator does not hold for the inverse problems considered
	in the current paper, so we focus instead on classical optimization schemes such as
	the ones described above.
	
	\subsection{Deep unfolding networks}
	More recently, deep unfolding networks (DUNs)~\cite{diamond_unrolled_2018,zhang_deep_2020} have been proposed for specific image restoration tasks. These approaches (like unrolling algorithms in general~\cite{Monga2021}) incorporate some advantages of both learning-based and model-based methods. Compared to standard learning-based methods, unrolled algorithms can be trained correctly on much smaller datasets~\cite{Gilton2019}. Compared to PnP methods, DUNs usually yield better results in fewer iterations, as they translate the truncated unfolded optimization into an end-to-end training of a deep network. An additional advantage of the latter is that the manual setting of optimization hyperparameters can be avoided in the unfolded scheme. In contrast, PnP methods remain more flexible and versatile, as DUNs need a separate training for each restoration task.
	
	\subsection{Video restoration with Plug-and-Play methods}
	To the best of our knowledge, this paper is the first one describing the use of neural networks in a PnP method for video restoration problems. In~\cite{yuan_plug-and-play_2021}, the video denoiser FastDVDnet~\cite{tassano_fastdvdnet_2020} is used in a PnP scheme, but to solve a specific problem of snapshot compressive imaging where the observation is a single frame. The work~\cite{khorasanighassab_plug-and-play_2021} uses the PnP-ADMM algorithm for video super-resolution, but employs the patch-based single-image denoiser BM3D~\cite{dabov_image_2007}.
	In the rest of this paper, we explore the use of video denoisers as regularizers within a PnP scheme and how they compare to the use of image denoisers on each frame separately. Section~\ref{sec:vpnp} describes our PnP method and how it integrates image and video denoisers. In  Section~\ref{sec:experiments}, we perform an extensive experimental evaluation of these algorithms on video deblurring, super-resolution and interpolation of random missing pixels. Concluding remarks and opportunities for future research are presented in Section~\ref{sec:conclusion}.
	
	\section{Video Plug-and-Play}\label{sec:vpnp}
	
	\subsection{PnP-ADMM.} Let us start by recalling the principle of the ADMM optimization method. Suppose that we want to minimize~\eqref{eq:energy_minimization}. We start by defining the augmented Lagrangian
	{\small \begin{equation}\label{eq:lagrangian}
			L_{\varepsilon}(\mathbf{x}, \mathbf{z}, \mathbf{v}) =
			\frac 1 \alpha \underbrace{\frac{1}{2 \sigma_{\mathbf{n}}^{2}}\|\mathbf{y}-\mathcal{A}(\mathbf{x})\|_{2}^{2}}_{F(\mathbf{x}, \mathbf{y})} + \mathcal{R}(\mathbf{z})+\frac{1}{2\varepsilon}\|\mathbf{x}-\mathbf{z}\|_{2}^{2} + \mathbf{v}^T(\mathbf{x}-\mathbf{z})
	\end{equation}}
	that we wish to minimize in $(\mathbf{x}, \mathbf{z})$ and maximize in $\mathbf{v}$. When  $\varepsilon \rightarrow 0$, the solutions $(\mathbf{x}, \mathbf{z}, \mathbf{v})$ satisfy $\mathbf{x} - \mathbf{z} \rightarrow \mathbf{0}$, and thus give us solutions of~\eqref{eq:energy_minimization}. The previous formulation is optimized in an alternating fashion by a scheme of the type (setting $\mathbf{u} = \varepsilon \mathbf{v}$)
	\begin{equation}
		\begin{aligned}
			\mathbf{x}_{k+1} & \leftarrow \underset{\mathbf{x}}{\operatorname{argmin}}L_{\varepsilon}(\mathbf{x}, \mathbf{z}_{k}, \mathbf{u}_{k}/\varepsilon) =  \operatorname{prox}_{\frac{\varepsilon}{\alpha} F(.,\mathbf{y})} (\mathbf{z}_{k}- \mathbf{u}_{k}) \\
			\mathbf{z}_{k+1} & \leftarrow \underset{\mathbf{z}}{\operatorname{argmin}}L_{\varepsilon}(\mathbf{x}_{k+1}, \mathbf{z}, \mathbf{u}_{k}/\varepsilon) = {\operatorname{prox}_{\varepsilon \mathcal{R}}} (\mathbf{x}_{k+1} +\mathbf{u}_{k}) \\
			\mathbf{u}_{k+1} & \leftarrow \mathbf{u}_{k}+\mathbf{x}_{k+1}-\mathbf{z}_{k+1},
		\end{aligned}
	\end{equation}
	where, if $\mathcal{A}$ is a linear operator represented by a matrix $A$, 
	\begin{equation}
	\label{eq:prox_data}
	    \operatorname{prox}_{\tau F(.,\mathbf{y})} (\mathbf{z}) = \left(\frac \tau {\sigma_n^2} A^* A + I_d\right)^{-1}\left(\frac \tau{\sigma_n^2} A^* \mathbf{y} + \mathbf{z}\right),
	\end{equation}
	with $I_d$ the identity matrix in dimension $d$ and $A^*$ the adjoint matrix of $A$.
	
	Consider that we know how to build a denoiser $\mathcal{D}_\varepsilon$ which can be expressed as the MAP estimator for a denoising problem (for an i.i.d. Gaussian noise of variance $\varepsilon$) with log-prior $\mathcal{R}$. By definition of the MAP, we have exactly $\mathcal{D}_\varepsilon = \operatorname{prox}_{\varepsilon  \mathcal{R}}$, so we can directly "plug" this denoiser in the previous optimization scheme. In practice, these PnP schemes are successfully used even with denoisers that do not satisfy this property, and the study of their convergence is a very active research field~\cite{chan_plug-and-play_2016, ryu_plug-and-play_2019, xu_provable_2020, laumont_maximum--posteriori_2021, hurault_gradient_2022, Hurault2022b}. The corresponding PnP-ADMM algorithm is summarized in Alg.~\ref{alg:pnp_admm}.
	\begin{algorithm}%[!t]
		\caption{PnP-ADMM scheme}
		\DontPrintSemicolon
		\textbf{Require:} $\mathbf{x}_{0} \in \mathbb{R}^{d}, \mathbf{y} \in \mathbb{R}^{m}, K \in \mathbb{N}^{\star}, \varepsilon>0, \alpha>0$ \\
		\textbf{Initialization:} Set $\mathbf{z}_{0}=\mathbf{x}_{0}$, and $\mathbf{u}_{k}=\mathbf{0}$\\
		\For{$k\in \left\{0,\ldots,K-1\right\}$}{
			$\begin{aligned}
				\mathbf{x}_{k+1} & \leftarrow \operatorname{prox}_{\frac{\varepsilon}{\alpha} F(.,\mathbf{y})}\left(\mathbf{z}_{k}-\mathbf{u}_{k}\right) \\
				\mathbf{z}_{k+1} & \leftarrow \mathcal{D}_\varepsilon\left(\mathbf{x}_{k+1}+\mathbf{u}_{k}\right) \\
				\mathbf{u}_{k+1} & \leftarrow \mathbf{u}_{k}+\left(\mathbf{x}_{k+1}-\mathbf{z}_{k+1}\right)
			\end{aligned}$
		}
		\textbf{return} $\mathbf{x}_{K}$
		\label{alg:pnp_admm}
	\end{algorithm}
	
	\subsection{Case of video}
	In the case of inverse problems on digital videos, we propose in this article to use the previous scheme directly on the whole video $\mathbf{x}$, which means that the proximal an denoising steps in the PnP-ADMM scheme are directly applied on the whole sequence. This allows considering cases where the operator $\mathcal{A}$ cannot be written in a separable way over all images of the sequence (in the case of temporal blur, for example). This also allows the use of networks specifically designed for video denoising, such as~\cite{tassano_fastdvdnet_2020}.
	
	Note that if one uses a single-frame denoiser (applied separately on each frame of the video), and if the degradation operator $\mathcal{A}$ is separable ({\em i.e.} can be written as a degradation on each frame of $\mathbf{x}$ separately), then the iterative ADMM scheme working on the whole video is equivalent to a succession of iterative ADMM schemes applied on each frame of the video. This is obviously no longer the case with video specific denoisers that are applied to the video in a non-separable way.
	
	\subsection{Performance of deep PnP Gaussian denoisers} \label{denoising}
	
	Here we focus on DRUNet~\cite{zhang_plug-and-play_2021} and FastDVDnet~\cite{tassano_fastdvdnet_2020}, two state-of-the-art networks for Gaussian denoising. DRUNet is a single-image network designed specifically to be integrated within a PnP approach for image restoration. FastDVDnet is a network designed for video denoising: it makes use of the additional information contained in neighboring images to provide a better and more temporally stable denoised estimate, with no explicit image alignment. Both networks use U-Net autoencoders~\cite{ronneberger_u-net_2015}: FastDVDnet combines two small U-Net blocks with residual connections and batch-norm in a cascaded architecture; DRUNet has a single deeper block with more subsampling steps and replaces the standard convolution layers with ResBlocks~\cite{lim_enhanced_2017}.
	
	We use these networks (and their weights) as provided by the authors of the original publications, without re-training them. Before studying the performance of these networks in PnP restoration, it is interesting to evaluate their denoising performance. It seems reasonable to think that the denoising performance of the network has an impact on the maximum achievable performance in PnP restoration~\cite{zhang_plug-and-play_2021}. DRUNet denoises all the frames of the video separately. To produce a denoised version of the image at time $t$, FastDVDnet uses the images at times $t_{-2}$, $t_{-1}$, $t$, $t_{+1}$ and $t_{+2}$. To ensure the output video retains the same size, the noisy input video is increased by 4 frames by mirroring: $\mathbf{x}= \{x_0, \dots, x_{N-1}\} \rightarrow \mathbf{x}^\prime= \{x_2, x_1, x_0, \dots, x_{N-1}, x_{N-2}, x_{N-3}\}.$
	The range of noise levels seen in training is $\sigma \in \left[5/255, 55/255\right]$ for FastDVDnet and $\sigma \in \left[0, 50/255\right]$ for DRUNet.
	
	We evaluate the video denoising performance of both networks on the test set of the DAVIS-2017 dataset~\cite{Pont-Tuset_arXiv_2017}, in its 480p version. This dataset consists in 30 video sequences of varying length. Neither DRUNet nor FastDVDnet have seen these sequences in training. The results can be seen in Table~\ref{table:denoise_davis}.
	
	\begin{table*}
		\caption{\label{table:denoise_davis}Denoising: PSNR/SSIM on DAVIS-2017-test-480p~\cite{Pont-Tuset_arXiv_2017} (8 CPU AMD 7F52 / 1 NVIDIA Tesla T4 / 16GB RAM)}
		\centering
		{\small
			\begin{tabular}{c| *{3}{c} c | c| c}
				\toprule
				Network   & $\sigma_{\mathbf{n}}=10/255$ & $\sigma_{\mathbf{n}}=25/255$ & $\sigma_{\mathbf{n}}=50/255$ & $\sigma_{\mathbf{n}}=100/255$& \# params       & exec. time (s / image) \\
				\midrule
				noisy     & 28.13/0.634 & 20.17/0.314 & 14.15/0.146 & 08.13/0.053 &           &     \\
				DRUNet     & 38.90/0.967 & 34.40/0.921 & 31.27/0.861 & \textbf{28.32/0.781} & 32.6410M  & 0.48\\	
				FastDVDnet & \textbf{39.20/0.969} & \textbf{35.05/0.931} & \textbf{31.97/0.878} & 26.84/0.655 & 2.4791M   & 0.23\\		
				\bottomrule
		\end{tabular}}
	\end{table*}
	FastDVDnet performs slightly better at most noise levels, even though it has around 13 times less parameters than DRUNet and takes about half the time to denoise one video frame. DRUNet, however, is the network whose performance degrades the least at $\sigma_{\mathbf{n}}=100/255$, a noise level not seen in training for both networks. Like FastDVDnet, the convolution layers of DRUNet are without biases; but DRUNet has no batch-norm layers, while FastDVDnet does (with biases in them). These results are consistent with those of ~\cite{mohan_robust_2020}, where it was shown that unbiased networks generalize better at noise levels outside the interval seen in training.
	
	\section{Experiments}\label{sec:experiments}
	
	We are now interested in the performance of our PnP-ADMM method using the networks of Section~\ref{denoising} for three video restoration problems: non-blind deblurring, super-resolution and interpolation of random missing pixels. These results aim at presenting the viability of our method, and at evaluating the impact of the chosen denoiser on the final result; other use cases, along with comparisons to methods specifically designed for such problems, will be the topic of future work.
	
	In the following experiments, we restrict each sequence to its first 30 frames to harmonize computation times and complexity per video. For each restoration problem, we find optimal values of the PnP parameters $(\varepsilon, \alpha, K)$ for a given denoiser with a grid search on a subset of the test data. Please refer to the Supplementary Material for more details on this procedure.
	
	\subsection{Non-blind video deblurring}\label{sec_deblurring}
	
	Our first use case is relatively straightforward: we generate blurry videos by convolving each frame by a known 2D kernel and adding noise of standard deviation $\sigma_{\mathbf{n}}$ to the result. For the sake of simplicity, we assume a cyclic convolution operator $\mathbf{k}$. The convolution with $\mathbf{k}$ can also be represented by a multiplication with a block circulant matrix $H$ with circulant blocks.
	%\begin{equation} \label{eq:forward_deblur}
	%	\mathbf{y} = \mathbf{k} \otimes %\mathbf{x} + \mathbf{n}.
	%\end{equation}
	 The degradation model becomes 
		\begin{equation} \label{eq:forward_deblur}
		\mathbf{y} = \mathbf{k} \otimes \mathbf{x} + \mathbf{n} = H \mathbf{x} + \mathbf{n}.
	\end{equation}

	In order to simulate blur which varies throughout the video (\emph{e.g.} blur caused by camera shake due to random hand tremor), we randomly sample a kernel among the 8 samples used in~\cite{levin_understanding_2009} for each frame. While this type of non-blind deconvolution problem can also be solved on a frame-by-frame basis, it is still relevant for digital video.
	
	% 	def prox_deblurring_torch(x, im_b, h, sigma, gamma):
	%     """
	%     Proximal Operator for Gaussian deblurring:
	%     f(x) = || A.x - im_b ||^2 / (2 sigma^2) 
	%     avec A.x = h*x
	%     prox_{gamma f} (x[i]) = 1/(1+gamma/sigma^2*h_fft*hc_fft)*(gamma/sigma^2 A.T y[i] +x[i]) 
	%     Parameters:
	%         :x - the argument to the proximal operator.
	%         :im_b - the noisy observation.
	%         :h_fft - FFT2 of the blurring kernel.
	%         :sigma - the standard deviation of the gaussian noise in y.
	%         :gamma - the regularization parameter.
	%     """
	
	%     s2 = max(0.255/255., sigma)**2  # trick to avoid division by 0 in noiseless case
	%     a = gamma/s2
	
	%     h_fft = torch.fft.fft2(h)
	%     hc_fft = torch.conj(h_fft)
	
	%     X = a*torch.real(torch.fft.ifft2(hc_fft*torch.fft.fft2(im_b))) + x
	
	%     return torch.real(torch.fft.ifft2(torch.fft.fft2(X)/(a*h_fft*hc_fft + 1))).float()
	
Applying $H$ can be done very simply in the Fourier domain, writing $ \mathcal{F}(H \mathbf{x})  = \mathcal{F}(\mathbf{k}) \mathcal{F}( \mathbf{x})$,
where $\mathbf{k}$ denotes the 2D blur kernel,
 $\mathcal{F}(.)$ denotes the 2D Fourier transform, $\overline{\mathcal{F}(.)}$ its conjugate and $\mathcal{F}^{-1}(.)$ its inverse.
	
	 Using~\eqref{eq:prox_data} with $A = H$ and the fact that the operators $H$ and  $H^*H$ are diagonal in the Fourier domain, we see that the proximal operator of the data fidelity term of~\eqref{eq:energy_minimization} can be written as
	\begin{align}
	\operatorname{prox}_{\tau F(.,\mathbf{y})} (\mathbf{z}) &= \left(\frac \tau {\sigma_n^2} H^* H + I_d\right)^{-1}\left(\frac \tau{\sigma_n^2} H^* \mathbf{y} + \mathbf{z}\right)\\
	&=\mathcal{F}^{-1}\left( \frac{  \frac{\tau}{ \sigma_{\mathbf{n}}^{2}}  \overline{\mathcal{F}(\mathbf{k})} \mathcal{F}(\mathbf{y}) + \mathcal{F}(\mathbf{z})  }{\frac{\tau}{ \sigma_{\mathbf{n}}^{2}} \overline{\mathcal{F}(\mathbf{k})} \mathcal{F}(\mathbf{k})+ 1} \right).
	\label{eq:prox_deblurring}
	\end{align}

	%
	 %$\mathbf{B}$ is the matrix that yields $\mathbf{B} \mathbf{x} \equiv \mathbf{k} \otimes \mathbf{x}$

\begin{figure*}%[t]
	\centering
	\begin{subfigure}[t]{\linewidth} % masked	
		\centering
		\begin{tikzpicture}[spy using outlines={every spy on node/.append style={smallwindow}}]
			\node [anchor=north] (FigA) at (0,0) {\includegraphics[width=.31\linewidth]{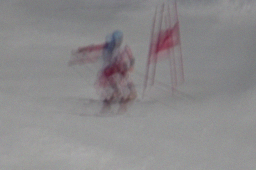}};
			\node [anchor=north] (FigB) at (0.33\linewidth,0) {\includegraphics[width=.31\linewidth]{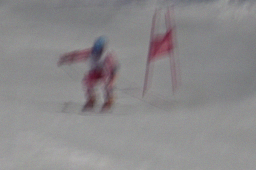}};
			\node [anchor=north] (FigC) at (0.66\linewidth,0) {\includegraphics[width=.31\linewidth]{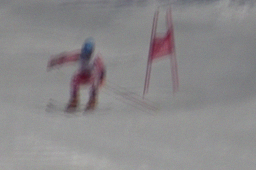}};
		\end{tikzpicture}	
		\subcaption{Observation PSNR = 28.34 dB}  %video: giant-slalom       PSNR/SSIM noisy: 28.34/0.7571, PSNR/SSIM out: 39.06/0.9493
	\end{subfigure}
	\begin{subfigure}[t]{\linewidth} % DRUNet
		\centering
		\begin{tikzpicture}[spy using outlines={every spy on node/.append style={smallwindow}}]
			\node [anchor=north] (FigA) at (0,0) {\includegraphics[width=.31\linewidth]{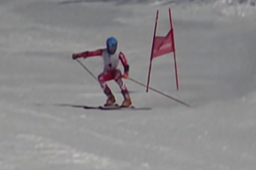}};
			\spy [closeup,magnification=2] on ($(FigA) + (-1.5,-1.54)$) 
			in node[largewindow,anchor=south east] at ($(FigA.south east) + (0.00, 0.00)$);
			\spy [closeup,magnification=2] on ($(FigA) + (0.8, 0.8)$) 
			in node[largewindow,anchor=north east] at ($(FigA.north east) + (-0.00, +0.00)$);
			\node [anchor=north] (FigB) at (0.33\linewidth,0) {\includegraphics[width=.31\linewidth]{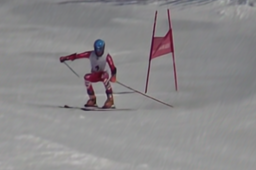}};
			\spy [closeup,magnification=2] on ($(FigB) + (-1.5,-1.54)$) 
			in node[largewindow,anchor=south east] at ($(FigB.south east) + (0.00, 0.00)$);
			\spy [closeup,magnification=2] on ($(FigB) + (0.8, 0.8)$) 
			in node[largewindow,anchor=north east] at ($(FigB.north east) + (-0.00, +0.00)$);
			\node [anchor=north] (FigC) at (0.66\linewidth,0) {\includegraphics[width=.31\linewidth]{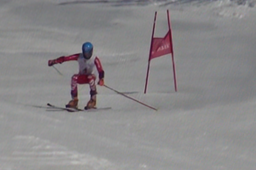}};
			\spy [closeup,magnification=2] on ($(FigC) + (-1.5,-1.54)$) 
			in node[largewindow,anchor=south east] at ($(FigC.south east) + (0.00, 0.00)$);
			\spy [closeup,magnification=2] on ($(FigC) + (0.8, 0.8)$) 
			in node[largewindow,anchor=north east] at ($(FigC.north east) + (-0.00, +0.00)$);
		\end{tikzpicture}
		\subcaption{Ours - DRUNet ($\sqrt{\varepsilon}=20/255, \alpha=1.5$) PSNR = 39.06 dB}
	\end{subfigure}
	\begin{subfigure}[t]{\linewidth} % FastDVDnet
		\centering
		\begin{tikzpicture}[spy using outlines={every spy on node/.append style={smallwindow}}]
			\node [anchor=north] (FigA) at (0,0) {\includegraphics[width=.31\linewidth]{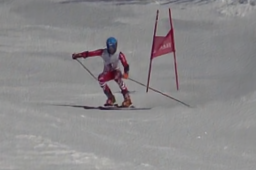}};
			\spy [closeup,magnification=2] on ($(FigA) + (-1.5,-1.54)$) 
			in node[largewindow,anchor=south east] at ($(FigA.south east) + (0.00, 0.00)$);
			\spy [closeup,magnification=2] on ($(FigA) + (0.8, 0.8)$) 
			in node[largewindow,anchor=north east] at ($(FigA.north east) + (-0.00, +0.00)$);
			\node [anchor=north] (FigB) at (0.33\linewidth,0) {\includegraphics[width=.31\linewidth]{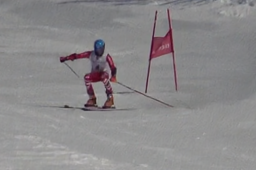}};
			\spy [closeup,magnification=2] on ($(FigB) + (-1.5,-1.54)$) 
			in node[largewindow,anchor=south east] at ($(FigB.south east) + (0.00, 0.00)$);
			\spy [closeup,magnification=2] on ($(FigB) + (0.8, 0.8)$) 
			in node[largewindow,anchor=north east] at ($(FigB.north east) + (-0.00, +0.00)$);
			\node [anchor=north] (FigC) at (0.66\linewidth,0) {\includegraphics[width=.31\linewidth]{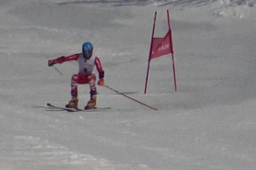}};
			\spy [closeup,magnification=2] on ($(FigC) + (-1.5,-1.54)$) 
			in node[largewindow,anchor=south east] at ($(FigC.south east) + (0.00, 0.00)$);
			\spy [closeup,magnification=2] on ($(FigC) + (0.8, 0.8)$) 
			in node[largewindow,anchor=north east] at ($(FigC.north east) + (-0.00, +0.00)$);
		\end{tikzpicture}
		\subcaption{Ours - FastDVDnet ($\sqrt{\varepsilon}=20/255, \alpha=1.0$) PSNR = 40.38 dB}  %PSNR/SSIM noisy: 27.92/0.7480, PSNR/SSIM out: 40.38/0.955
	\end{subfigure}
	\begin{subfigure}[t]{\linewidth} % DPIR
		\centering
		\begin{tikzpicture}[spy using outlines={every spy on node/.append style={smallwindow}}]
			\node [anchor=north] (FigA) at (0,0) {\includegraphics[width=.31\linewidth]{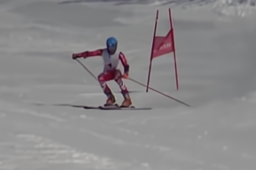}};
			\spy [closeup,magnification=2] on ($(FigA) + (-1.5,-1.54)$) 
			in node[largewindow,anchor=south east] at ($(FigA.south east) + (0.00, 0.00)$);
			\spy [closeup,magnification=2] on ($(FigA) + (0.8, 0.8)$) 
			in node[largewindow,anchor=north east] at ($(FigA.north east) + (-0.00, +0.00)$);
			\node [anchor=north] (FigB) at (0.33\linewidth,0) {\includegraphics[width=.31\linewidth]{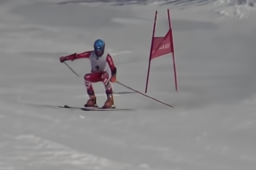}};
			\spy [closeup,magnification=2] on ($(FigB) + (-1.5,-1.54)$) 
			in node[largewindow,anchor=south east] at ($(FigB.south east) + (0.00, 0.00)$);
			\spy [closeup,magnification=2] on ($(FigB) + (0.8, 0.8)$) 
			in node[largewindow,anchor=north east] at ($(FigB.north east) + (-0.00, +0.00)$);
			\node [anchor=north] (FigC) at (0.66\linewidth,0) {\includegraphics[width=.31\linewidth]{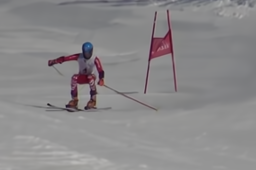}};
			\spy [closeup,magnification=2] on ($(FigC) + (-1.5,-1.54)$) 
			in node[largewindow,anchor=south east] at ($(FigC.south east) + (0.00, 0.00)$);
			\spy [closeup,magnification=2] on ($(FigC) + (0.8, 0.8)$) 
			in node[largewindow,anchor=north east] at ($(FigC.north east) + (-0.00, +0.00)$);
		\end{tikzpicture}
		\subcaption{DPIR~\cite{zhang_plug-and-play_2021} PSNR = 39.77 dB}  %video: giant-slalom       PSNR/SSIM noisy: 28.34/0.7571, PSNR/SSIM out: 39.77/0.9619
	\end{subfigure}	
	\caption{\label{fig:deblurring}Video deblurring ($\sigma_{\mathbf{n}}=2.55$) on \texttt{giant-slalom} of DAVIS-2017-test-480p~\cite{Pont-Tuset_arXiv_2017}. Ours - DRUNet and FastDVDnet are using optimal values of $(\varepsilon, \alpha)$. While all methods perform well, Ours - FastDVDnet restores more image detail and its results are more temporally stable, whereas DPIR yields smoother frames with somewhat less high frequency content.}
    \end{figure*}

	\noindent\textbf{Results.} The quantitative results of our PnP-ADMM method for video deblurring with the 8 randomized kernels of~\cite{levin_understanding_2009} on DAVIS-2017-test-480p~\cite{Pont-Tuset_arXiv_2017} for $\sigma_{\mathbf{n}} \in \left\{2.55, 7.65\right\}/255$ are presented in Table~\ref{table:deblur_davis}.
	For reference, the table also includes the comparison of our method to DPIR~\cite{zhang_plug-and-play_2021} on the same dataset. It is a state-of-the-art PnP image restoration algorithm, which uses DRUNet in PnP-HQS with an annealing strategy on the noise level of their denoiser across iterations. In that case, we perform PnP restoration of each video frame sequentially, using the parameter values recommended by the authors for deblurring: $(\sigma_{1},\sigma_{K}, \lambda, K) = (49, \sigma_{\mathbf{n}} \times 255, 0.23, 8)$ using the notations of~\cite{zhang_plug-and-play_2021}. We turn off periodical geometric self-ensemble in DPIR (which consists in rotating / flipping $\mathbf{x}$ and $\mathbf{z}$ at each iteration with a period of 8 iterations to improve the performance of the denoising step), as it could also be implemented in our method but only yields marginal improvements and is not necessarily relevant to the comparisons of this article.
	\begin{table}[t]
		\caption{\label{table:deblur_davis}Video deblurring: PSNR/SSIM on DAVIS-2017-test-480p~\cite{Pont-Tuset_arXiv_2017} (each video limited to its 30 first frames)}
		\centering
		{\small
			\begin{tabular}{c| c c}
				\toprule
				Method           & $\sigma_{\mathbf{n}}=2.55/255$  & $\sigma_{\mathbf{n}}=7.65/255$ \\
				\midrule
				blurred          & 22.75/0.5994 & 12.66/0.533 \\
				Ours - DRUNet     & 36.20/0.938 & 32.07/0.856 \\
				Ours - FastDVDnet & \textbf{37.11}/0.949 & \textbf{33.29/0.893} \\
				DPIR~\cite{zhang_plug-and-play_2021} & 36.97/\textbf{0.950} & 32.92/0.890 \\
				\bottomrule
		\end{tabular}}
	\end{table}
	With our video PnP-ADMM method, using FastDVDnet instead of DRUNet yields a higher PSNR/SSIM in almost all of the 30 videos observed at both noise levels. The performance gap is smaller between PnP-ADMM using FastDVDnet and DPIR (which also uses DRUNet), which suggests that using a different optimization algorithm (and in turn a different set of parameters) for video Plug-And-Play could lead to different results with identical denoisers. Figure~\ref{fig:deblurring} showcases a visual comparison of results.
    It can be observed that FastDVDnet restores more image detail and its results are more temporally stable, whereas DPIR yields smoother frames with somewhat less high frequency content.

	\subsection{Video super-resolution}\label{sec_sr}
	
	For video super-resolution, we use the classical degradation model where the low resolution video is obtained by downsampling the original high resolution video by a certain scale factor $s$ in each dimension after convolution by a known anti-aliasing kernel $\mathbf{k}$. Writing $H$ the $d\times d$ matrix representing the cyclic convolution operator and $S$ the $m\times d$ downsampling matrix, the degradation model can be written 
	\begin{equation}\label{eq:forward_sr}
		\mathbf{y} = SH \mathbf{x} + \mathbf{n}.
	\end{equation}
	Replacing $A$ by $SH$ in \eqref{eq:prox_data},  we deduce that the proximal operator of the data fidelity term can be written in this case
	\begin{equation}
	\operatorname{prox}_{\tau F(.,\mathbf{y})} (\mathbf{z}) = \left(\frac \tau {\sigma_n^2} H^*S^* S H + I_d\right)^{-1}\left(\frac \tau{\sigma_n^2} H^*S^* \mathbf{y} + \mathbf{z}\right). 
		\end{equation}
	Using this formulation directly is not possible since it requires to invert a huge $d\times d$ matrix, which is not diagonal in the frequency domain.
	Instead, as done in~\cite{zhang_plug-and-play_2021, hurault_gradient_2022}, we use the closed-form  expression proposed in~\cite{zhao_fast_2016}. 
	First, following the notations of~\cite{hurault_gradient_2022}, we write $\hat{z}_{\tau} =\frac \tau{\sigma_n^2} H^*S^* \mathbf{y} + \mathbf{z}$ and observe that
	\begin{align}
	\operatorname{prox}_{\tau F(.,\mathbf{y})} (\mathbf{z})& = \left(\frac \tau {\sigma_n^2} H^*S^* S H + I_d\right)^{-1} \hat{z}_{\tau}\\ =& \hat{z}_{\tau} - \frac \tau {\sigma_n^2} H^*S^* \left(\frac \tau {\sigma_n^2}  S H H^*S^*+ I_m\right)^{-1} S H \hat{z}_{\tau}.
	\end{align}
	The $m\times m$ matrix $S H H^*S^* + I_m$ can be inverted much more easily in the Fourier domain. Following again the notations of~\cite{hurault_gradient_2022}, we write  $\Lambda = \mathrm{diag}(\mathcal{F}(\mathbf{k}))$ the $d\times d$ diagonal matrix containing the Fourier transform of the convolution kernel $\mathbf{k}$ on the diagonal. This  matrix can also be written as a block diagonal matrix    $\Lambda=\operatorname{diag}\left(\Lambda_{1}, \ldots, \Lambda_{s^{2}}\right)$, with blocks $\Lambda_{k}$ (also diagonal) of size $m\times m$. Now, writing  $\underline{\Lambda}=\left[\Lambda_{1}, \ldots, \Lambda_{s^{2}}\right] \in \mathbb{R}^{m \times d}$, it follows easily  that the operator $SH$  corresponds to $\frac 1 s \underline{\Lambda}$ in the Fourier domain, {\em i.e.} $\mathcal{F}(SH\mathbf{z}) = \frac 1 s \underline{\Lambda} \mathcal{F}(\mathbf{z})$ for all $\mathbf{z} \in \mathbb{R}^d$. Finally, the proximal operator of the data term can be computed explicitly  as 
	{\small
	\begin{equation}
			\operatorname{prox}_{\tau F(.,\mathbf{y})}(\mathbf{z})=\hat{z}_{\tau}-\frac{\tau}{\sigma_n^2s^{2}} \mathcal{F}^{-1}\left( \underline{\Lambda}^{*}\left(\frac{\tau}{\sigma_n^2s^{2}} \underline{\Lambda \Lambda^{*}} + I_m\right)^{-1} \underline{\Lambda} \mathcal{F}( \hat{z}_{\tau})\right).
	\end{equation}}
	
	 %and $\underline{\Lambda}=\left[\Lambda_{1}, \ldots, \Lambda_{s^{2}}\right] \in \mathbb{R}^{m \times n}$, with $\Lambda=\operatorname{diag}\left(\Lambda_{1}, \ldots, \Lambda_{s^{2}}\right)$ a blockdiagonal decomposition according to a $s \times s$ paving of the Fourier domain.

	%
	%\begin{equation}\label{eq:forward_sr}
	%	\mathbf{y} = (\mathbf{k} \otimes %\mathbf{x}) \downarrow_{s} + \mathbf{n}.
	%\end{equation}
	%
	
	%THIS IS THE ZHANG VERSION
	%\begin{equation}
	%	\mathbf{x}_{k}=\mathcal{F}^{-1}\left(\frac{1}{\alpha_{k}}\left(\mathbf{d}-\overline{\mathcal{F}(\mathbf{k})} \odot_{s} \frac{(\mathcal{F}(\mathbf{k}) \mathbf{d}) \Downarrow_{s}}{(\overline{\mathcal{F}(\mathbf{k})} \mathcal{F}(\mathbf{k})) \Downarrow_{s}+\alpha_{k}}\right)\right)
	%\end{equation}
	%"where $\mathbf{d}=\overline{\mathcal{F}(\mathbf{k})} \mathcal{F}\left(\mathbf{y} \uparrow_{s}\right)+\alpha_{k} \mathcal{F}\left(\mathbf{z}_{k-1}\right)$
	%and where $\odot_{s}$ denotes distinct block processing operator with element-wise multiplication, i.e., applying element-wise multiplication to the $s \times s$ distinct blocks of $\overline{\mathcal{F}(\mathbf{k})}$ and $\Downarrow_{s}$ denotes distinct block downsampler, i.e., averaging the $s \times s$ distinct blocks"
 	
	%NOTE TO JD / AA / MT: in the code, $\alpha_k$ corresponds to $\frac{\alpha \sigma_{\mathbf{n}}^{2}}{\varepsilon}$ using our notations

	%THIS IS THE HURAULT VERSION (they use the same code in practice)

\begin{figure*}%[!t]
	\centering
	\begin{subfigure}[t]{\linewidth} % bicubic
		\centering
		\begin{tikzpicture}[spy using outlines={every spy on node/.append style={smallwindow}}]
			\node [anchor=north] (FigA) at (0,0) {\includegraphics[width=.31\linewidth]{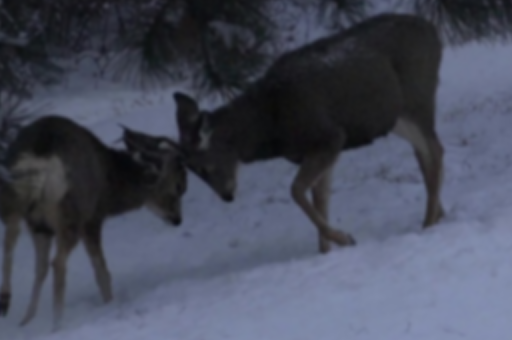}};
			\spy [closeup,magnification=3] on ($(FigA) + (-1.2, 1)$) 
			in node[largewindow,anchor=south west] at ($(FigA.south west) + (0.00, 0.00)$);
			\spy [closeup,magnification=2] on ($(FigA) + (0.6,-0.05)$) 
			in node[largewindow,anchor=south east] at ($(FigA.south east) + (-0.00, +0.00)$);
			\node [anchor=north] (FigB) at (0.33\linewidth,0) {\includegraphics[width=.31\linewidth]{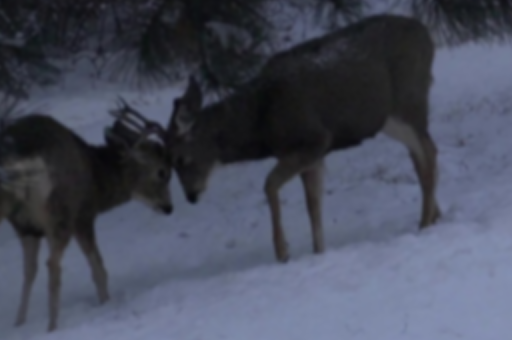}};
			\spy [closeup,magnification=3] on ($(FigB) + (-1.8,1.2)$) 
			in node[largewindow,anchor=south west] at ($(FigB.south west) + (0.00, 0.00)$);
			\spy [closeup,magnification=2] on ($(FigB) + (1, 0.6)$) 
			in node[largewindow,anchor=south east] at ($(FigB.south east) + (-0.00, +0.00)$);
			\node [anchor=north] (FigC) at (0.66\linewidth,0) {\includegraphics[width=.31\linewidth]{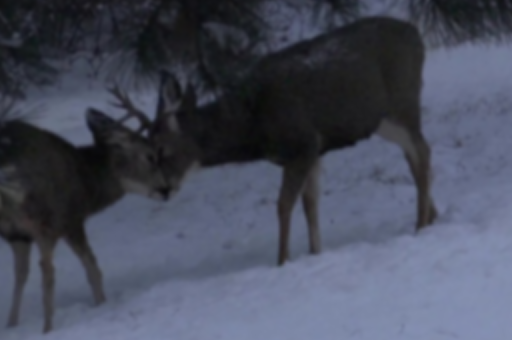}};
			\spy [closeup,magnification=2] on ($(FigC) + (-1.3,0)$) 
			in node[largewindow,anchor=south west] at ($(FigC.south west) + (0.00, 0.00)$);
			\spy [closeup,magnification=2] on ($(FigC) + (0.8,1.3)$) 
			in node[largewindow,anchor=south east] at ($(FigC.south east) + (-0.00, +0.00)$);
		\end{tikzpicture}	
		\subcaption{LR + bicubic PSNR = 34.17 dB}
	\end{subfigure}
	\begin{subfigure}[t]{\linewidth} % FastDVDnet
		\centering
		\begin{tikzpicture}[spy using outlines={every spy on node/.append style={smallwindow}}]
			\node [anchor=north] (FigA) at (0,0) {\includegraphics[width=.31\linewidth]{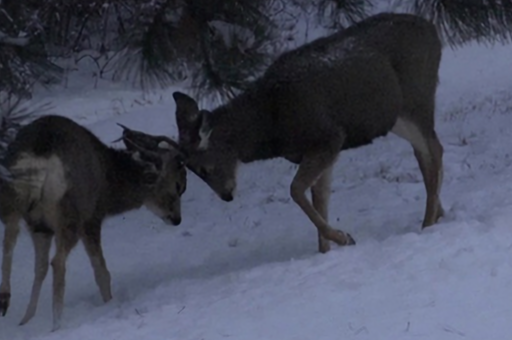}};
			\spy [closeup,magnification=3] on ($(FigA) + (-1.2, 1)$) 
			in node[largewindow,anchor=south west] at ($(FigA.south west) + (0.00, 0.00)$);
			\spy [closeup,magnification=2] on ($(FigA) + (0.6,-0.05)$) 
			in node[largewindow,anchor=south east] at ($(FigA.south east) + (-0.00, +0.00)$);
			\node [anchor=north] (FigB) at (0.33\linewidth,0) {\includegraphics[width=.31\linewidth]{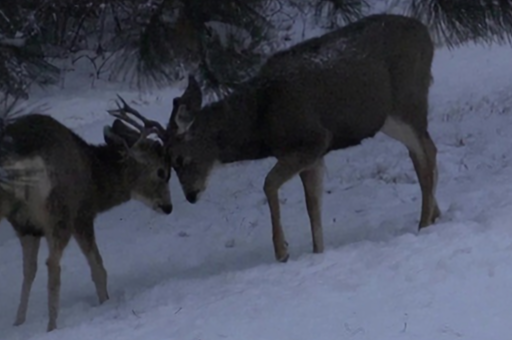}};
			\spy [closeup,magnification=3] on ($(FigB) + (-1.8,1.2)$) 
			in node[largewindow,anchor=south west] at ($(FigB.south west) + (0.00, 0.00)$);
			\spy [closeup,magnification=2] on ($(FigB) + (1, 0.6)$) 
			in node[largewindow,anchor=south east] at ($(FigB.south east) + (-0.00, +0.00)$);
			\node [anchor=north] (FigC) at (0.66\linewidth,0) {\includegraphics[width=.31\linewidth]{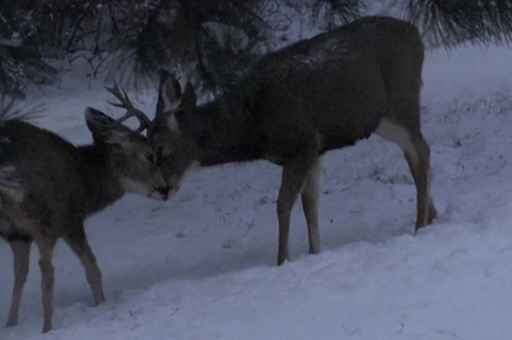}};
			\spy [closeup,magnification=2] on ($(FigC) + (-1.3,0)$) 
			in node[largewindow,anchor=south west] at ($(FigC.south west) + (0.00, 0.00)$);
			\spy [closeup,magnification=2] on ($(FigC) + (0.8,1.3)$) 
			in node[largewindow,anchor=south east] at ($(FigC.south east) + (-0.00, +0.00)$);
		\end{tikzpicture}
		\subcaption{Ours - FastDVDnet ($\sqrt{\varepsilon}=20/255, \alpha=0.025$) PSNR = 44.35 dB}  %video: deer               PSNR/SSIM noisy: 34.17/0.9253, PSNR/SSIM out: 44.35/0.9860
	\end{subfigure}
	\begin{subfigure}[t]{\linewidth} % DPIR
		\centering
		\begin{tikzpicture}[spy using outlines={every spy on node/.append style={smallwindow}}]
			\node [anchor=north] (FigA) at (0,0) {\includegraphics[width=.31\linewidth]{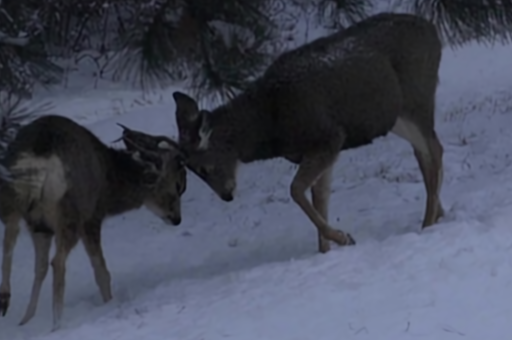}};
			\spy [closeup,magnification=3] on ($(FigA) + (-1.2, 1)$) 
			in node[largewindow,anchor=south west] at ($(FigA.south west) + (0.00, 0.00)$);
			\spy [closeup,magnification=2] on ($(FigA) + (0.6,-0.05)$) 
			in node[largewindow,anchor=south east] at ($(FigA.south east) + (-0.00, +0.00)$);
			\node [anchor=north] (FigB) at (0.33\linewidth,0) {\includegraphics[width=.31\linewidth]{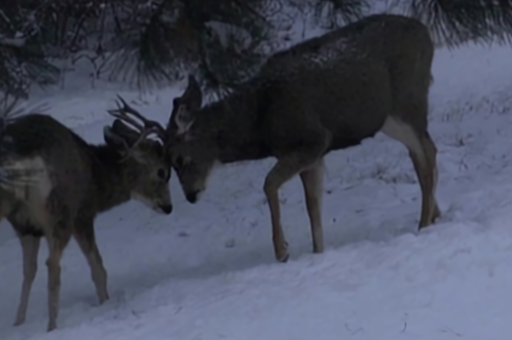}};
			\spy [closeup,magnification=3] on ($(FigB) + (-1.8,1.2)$) 
			in node[largewindow,anchor=south west] at ($(FigB.south west) + (0.00, 0.00)$);
			\spy [closeup,magnification=2] on ($(FigB) + (1, 0.6)$) 
			in node[largewindow,anchor=south east] at ($(FigB.south east) + (-0.00, +0.00)$);
			\node [anchor=north] (FigC) at (0.66\linewidth,0) {\includegraphics[width=.31\linewidth]{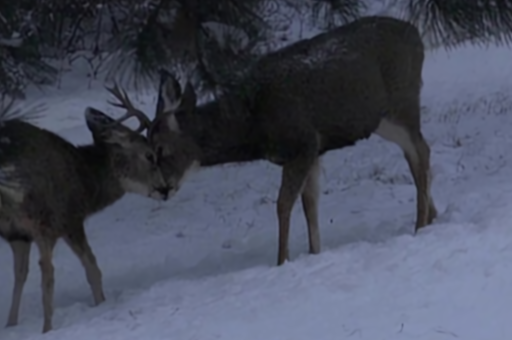}};
			\spy [closeup,magnification=2] on ($(FigC) + (-1.3,0)$) 
			in node[largewindow,anchor=south west] at ($(FigC.south west) + (0.00, 0.00)$);
			\spy [closeup,magnification=2] on ($(FigC) + (0.8,1.3)$) 
			in node[largewindow,anchor=south east] at ($(FigC.south east) + (-0.00, +0.00)$);
		\end{tikzpicture}
		\subcaption{DPIR~\cite{zhang_plug-and-play_2021} PSNR = 42.91 dB}  %video: deer               PSNR/SSIM noisy: 34.17/0.9253, PSNR/SSIM out: 42.91/0.9777
	\end{subfigure}	
	\caption{\label{fig:sr}Video SR ($\times2$/ Gauss. ($\sigma=1.6$), $\sigma_{\mathbf{n}}=0$) on the \texttt{deer} sequence of DAVIS-2017-test-480p~\cite{Pont-Tuset_arXiv_2017}. Our video PnP-ADMM method with FastDVDnet retains slightly more detail than DPIR (Ours - DRUNet is not shown as it is similar to Ours - FastDVDnet with 43.97 dB).}
	\end{figure*}

	\noindent\textbf{Results.} The quantitative results of our PnP-ADMM method for $\times2$ and $\times4$ super-resolution with two Gaussian kernels for $\sigma_{\mathbf{n}} \in \left\{0, 2.55, 7.65\right\}/255$ are presented in Table~\ref{table:sr_davis}.
	\begin{table*}%[!t]
		\caption{\label{table:sr_davis}Video super-resolution: PSNR/SSIM on DAVIS-2017-test-480p~\cite{Pont-Tuset_arXiv_2017} (each video limited to its 30 first frames)}
		\centering

% \usepackage{multirow}
    {\small
    \begin{tabular}{*{6}{c}}
    \toprule 
    s.f. / kernel &  & LR + bicubic & Ours - DRUNet & Ours - FastDVDnet                     & DPIR~\cite{zhang_plug-and-play_2021} \\
    \midrule
    \multirow{3}{*}{$\times2$/ Gauss. ($\sigma=1.6$)}  & $\sigma_{\mathbf{n}}=0$        & 27.76/0.816  & 36.47/0.955   & \textbf{36.86/0.958} & 35.62/0.945 \\   
    & $\sigma_{\mathbf{n}}=2.55/255$ & 27.52/0.784  & 32.12/0.874   & 32.37/0.874   & \textbf{32.56/0.895}\\   
    & $\sigma_{\mathbf{n}}=7.65/255$ & 26.06/0.615  & 29.86/0.806   & 30.25/0.817   & \textbf{30.59/0.845}\\
    \midrule
    \multirow{3}{*}{$\times4$ / Gauss. ($\sigma=1.6$)} & $\sigma_{\mathbf{n}}=0$        & 24.74/0.714  & 30.00/\textbf{0.844} & \textbf{30.12}/0.839 & 29.75/0.840 \\   
    & $\sigma_{\mathbf{n}}=2.55/255$ & 24.61/0.689  & 29.23/0.809   & \textbf{29.38}/0.805 & 29.31/\textbf{0.823}\\   
    & $\sigma_{\mathbf{n}}=7.65/255$ & 23.78/0.563  & 28.18/0.772   & \textbf{28.47}/0.780 & 28.29/\textbf{0.783}\\    
    \midrule
    \multirow{3}{*}{$\times4$ / Gauss. ($\sigma=3.2$)}                      & $\sigma_{\mathbf{n}}=0$        & 23.98/0.665  & 30.07/\textbf{0.844} & \textbf{30.15}/0.841 & 29.70/0.830 \\   
     & $\sigma_{\mathbf{n}}=2.55/255$ & 23.88/0.642  & 26.42/0.634   & 27.46/0.711   & \textbf{27.94/0.767}\\
     & $\sigma_{\mathbf{n}}=7.65/255$ & 23.17/0.518  & 26.37/0.698   & \textbf{26.54}/0.701 & 26.53/\textbf{0.713}                   \\ 
    \bottomrule
    \end{tabular}}

% 		{\small
% 			\begin{tabular}{c| *{2}{c c c|} c c c}
% 				\toprule
% 				s.f. / kernel    & \multicolumn{3}{c|}{$\times2$/ Gauss. ($\sigma=1.6$)} & \multicolumn{3}{c|}{$\times4$ / Gauss. ($\sigma=1.6$)} & \multicolumn{3}{c}{$\times4$ / Gauss. ($\sigma=3.2$)} \\
% 				%\cline{2-3} \cline{4-5} \cline{6-7}
% 				Method     & $\sigma_{\mathbf{n}}=0$     & $\sigma_{\mathbf{n}}=2.55/255$ & $\sigma_{\mathbf{n}}=7.65/255$ & $\sigma_{\mathbf{n}}=0$ & $\sigma_{\mathbf{n}}=2.55/255$ & $\sigma_{\mathbf{n}}=7.65/255$ & $\sigma_{\mathbf{n}}=0$     & $\sigma_{\mathbf{n}}=2.55/255$ & $\sigma_{\mathbf{n}}=7.65/255$ \\
% 				\midrule
% 				LR + bicubic       & 27.76/0.816  & 27.52/0.784 & 26.06/0.615 & 24.74/0.714 & 24.61/0.689 & 23.78/0.563 & 23.98/0.665  & 23.88/0.642 & 23.17/0.518\\
% 				Ours - DRUNet     & 36.47/0.955 & 32.12/0.874 & 29.86/0.806 & 30.00/\textbf{0.844} & 29.23/0.809 & 28.18/0.772 & 30.07/\textbf{0.844} & 26.42/0.634 & 26.37/0.698\\
% 				Ours - FastDVDnet & \textbf{36.86/0.958} & 32.37/0.874 & 30.25/0.817& \textbf{30.12}/0.839 & \textbf{29.38}/0.805 & \textbf{28.47}/0.780 & \textbf{30.15}/0.841 & 27.46/0.711 & \textbf{26.54}/0.701\\
% 				DPIR~\cite{zhang_plug-and-play_2021} & 35.62/0.945 & \textbf{32.56/0.895} & \textbf{30.59/0.845} & 29.75/0.840 & 29.31/\textbf{0.823} & 28.29/\textbf{0.783} & 29.70/0.830 & \textbf{27.94/0.767} & DPIR 26.53/\textbf{0.713}\\
% 				\bottomrule
% 		\end{tabular}}
	\end{table*}
% \begin{itemize}
% 	\small
% 	\item $\times2$ / Gauss. ($\sigma=1.6$) : LR+bicubic 26.06/0.615 PnP-V DRUNet ($\sqrt{\varepsilon} = 70/255$, $\alpha=1$, $K=10$) 29.86/0.806 PnP-V FastDVDnet ($\sqrt{\varepsilon} = 40/255$, $\alpha=0.5$, $K=20$) 30.25/0.817 DPIR \textbf{30.59/0.845}
% 	\item $\times4$ / Gauss. ($\sigma=1.6$) : LR+bicubic 23.78/0.563 PnP-V DRUNet ($\sqrt{\varepsilon} = 70/255$, $\alpha=1$, $K=10$) 28.18/0.772 PnP-V FastDVDnet ($\sqrt{\varepsilon} = 55/255$, $\alpha=0.75$, $K=20$) \textbf{28.47}/0.780 DPIR 28.29/\textbf{0.783}
% 	\item $\times4$ / Gauss. ($\sigma=3.2$) : LR+bicubic 23.17/0.518 PnP-V DRUNet ($\sqrt{\varepsilon} = 70/255$, $\alpha=0.5$, $K=10$) 26.37/0.698 PnP-V FastDVDnet ($\sqrt{\varepsilon} = 40/255$, $\alpha=0.25$, $K=20$) \textbf{26.54}/0.701 DPIR 26.53/\textbf{0.713}
% \end{itemize}
	Once again, we also include the comparison to DPIR~\cite{zhang_plug-and-play_2021}, with the parameter values recommended by the authors for super-resolution: $(\sigma_{1},\sigma_{K}, \lambda, K) = (49, \max(s, \sigma_{\mathbf{n}}\times 255), 0.23, 24)$ using the notations of~\cite{zhang_plug-and-play_2021}. As in Section \ref{sec_deblurring}, we turn off periodical geometric self-ensemble. Similarly to the previous restoration problem, PnP-ADMM with FastDVDnet produces the best results in most cases. An example is illustrated in Figure \ref{fig:sr}.
	
	\subsection{Video interpolation of random missing pixels} \label{sec_interpolation}

	We evaluate a final use case of our Plug-and-Play video restoration method: interpolation, which consists in estimating the values of hidden or missing pixels. In our experiments, we mask a proportion $\rho$ of the pixels in the video according to a random spatio-temporal pattern. This inverse problem can be seen as a special case of compressed sensing~\cite{donoho_compressed_2006}. The degradation model can be expressed as
	\begin{equation}\label{eq:forward_interp}
		\mathbf{y} = \mathbf{M} \odot \mathbf{x} + \mathbf{n}
	\end{equation}
	where $\mathbf{M}$ is random with a proportion $\rho$ of elements set to 0 and 1 elsewhere and $\odot$ denotes pixel-wise multiplication.
	The proximal operator of the data term for this type of problem can be written as
	\begin{equation}\label{eq:prox_interp}
		\operatorname{prox}_{\tau F(.,\mathbf{y})}(\mathbf{x}[i])=
		\begin{cases}
			\frac{\mathbf{x}[i] + \mathbf{y}[i] \frac{\tau}{\sigma_{\mathbf{n}}^{2}}}{1 + \frac{\tau}{\sigma_{\mathbf{n}}^{2}}} & \text{if $\mathbf{M}[i] = 1$}\\
			\mathbf{x}[i] & \text{if $\mathbf{M}[i] = 0$}
		\end{cases}
	\end{equation}
	where $\mathbf{x}[i]$ is the value of the pixel $i$ at a specific spatial location of a specific channel of a specific frame of the RGB video $\mathbf{x}$.

%\iota_{C}(x)= \begin{cases}0 & \text { if } x \in C \\ +\infty & \text { otherwise }\end{cases}

	In the case where $\sigma_{\mathbf{n}} = 0$, the data-fitting term takes the form of a hard constraint and its proximal operator admits a closed form that is independent of $\tau$ (and thus of $\alpha$ in our PnP-ADMM method since we have $\tau \equiv \varepsilon/\alpha$):
	
	\begin{equation}\label{eq:prox_interp_nonoise}
		\operatorname{prox}_{\tau F(.,\mathbf{y})} (\mathbf{x}) = (1 - \mathbf{M}) \odot \mathbf{x} + \mathbf{M} \odot \mathbf{y}.
	\end{equation}

\begin{figure*}[t]
		\centering
% 			\begin{subfigure}[t]{\linewidth} % clean
% 				\centering
% 				% 	\begin{subfigure}[t]{.24\linewidth}
% 				% 		\centering
% 				% 			\includegraphics[width=\linewidth]{images/horsejump-stick/horsejump-stick_clean/horsejump-stick_clean_025_cropped.png}
% 				% 	\end{subfigure}
% 					\begin{subfigure}[t]{.24\linewidth}
% 						\centering
% 							\includegraphics[width=\linewidth]{images/horsejump-stick/horsejump-stick_clean/horsejump-stick_clean_026_cropped.png}
% 					\end{subfigure}
% 					\begin{subfigure}[t]{.24\linewidth}
% 						\centering
% 							\includegraphics[width=\linewidth]{images/horsejump-stick/horsejump-stick_clean/horsejump-stick_clean_027_cropped.png}
% 					\end{subfigure}
% 					\begin{subfigure}[t]{.24\linewidth}
% 						\centering
% 							\includegraphics[width=\linewidth]{images/horsejump-stick/horsejump-stick_clean/horsejump-stick_clean_028_cropped.png}
% 					\end{subfigure}
% 					\begin{subfigure}[t]{.24\linewidth}
% 						\centering
% 							\includegraphics[width=\linewidth]{images/horsejump-stick/horsejump-stick_clean/horsejump-stick_clean_029_cropped.png}
% 					\end{subfigure}
% 				\subcaption{Unknown / ground truth}
% 			\end{subfigure}		
			%	
			\begin{subfigure}[t]{\linewidth} % masked
				\centering
%					\begin{subfigure}[t]{.24\linewidth}
%						\centering
%							\includegraphics[width=\linewidth]{images/horsejump-stick/horsejump-stick_masked_p0.9_sigma0.0/horsejump-stick_masked_025_cropped.png}
%					\end{subfigure}
					\begin{subfigure}[t]{.24\linewidth}
						\centering
							\includegraphics[width=\linewidth]{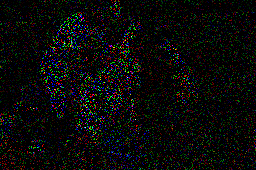}
					\end{subfigure}
					\begin{subfigure}[t]{.24\linewidth}
						\centering
							\includegraphics[width=\linewidth]{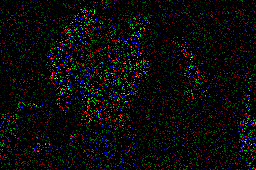}
					\end{subfigure}
					\begin{subfigure}[t]{.24\linewidth}
						\centering
							\includegraphics[width=\linewidth]{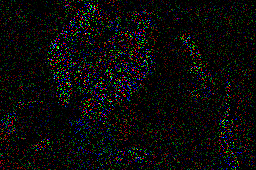}
					\end{subfigure}
					\begin{subfigure}[t]{.24\linewidth}
						\centering
							\includegraphics[width=\linewidth]{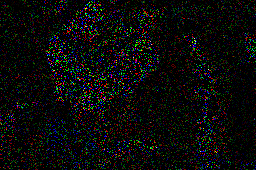}
					\end{subfigure}
				\subcaption{Observation PSNR = 8.23 dB}
			\end{subfigure}
			\begin{subfigure}[t]{\linewidth} % DRUNet
				\centering
%					\begin{subfigure}[t]{.24\linewidth}
%						\centering
%							\includegraphics[width=\linewidth]{images/horsejump-stick/horsejump-stick_masked_p0.9_sigma0.0_DRUNet_alpha2.75_s50_200iters_restored/horsejump-stick_restored_025_cropped.png}
%					\end{subfigure}
					\begin{subfigure}[t]{.24\linewidth}
						\centering
							\includegraphics[width=\linewidth]{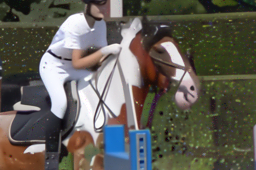}
					\end{subfigure}
					\begin{subfigure}[t]{.24\linewidth}
						\centering
							\includegraphics[width=\linewidth]{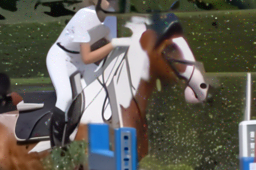}
					\end{subfigure}
					\begin{subfigure}[t]{.24\linewidth}
						\centering
							\includegraphics[width=\linewidth]{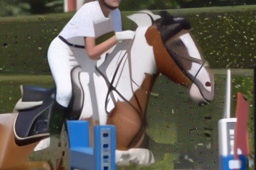}
					\end{subfigure}
					\begin{subfigure}[t]{.24\linewidth}
						\centering
							\includegraphics[width=\linewidth]{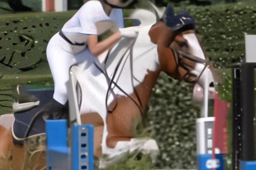}
					\end{subfigure}
				\subcaption{Ours - DRUNet ($\sqrt{\varepsilon}=50/255, \alpha=2.75$) PSNR = 26.18 dB}
			\end{subfigure}
			\begin{subfigure}[t]{\linewidth} % FastDVDnet
				\centering
%					\begin{subfigure}[t]{.24\linewidth}
%						\centering
%							\includegraphics[width=\linewidth]{images/horsejump-stick/horsejump-stick_masked_p0.9_sigma0.0_FastDVDnet_alpha2.25_s30_200iters_restored/horsejump-stick_restored_025_cropped.png}
%					\end{subfigure}
					\begin{subfigure}[t]{.24\linewidth}
						\centering
							\includegraphics[width=\linewidth]{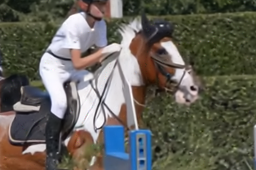}
					\end{subfigure}
					\begin{subfigure}[t]{.24\linewidth}
						\centering
							\includegraphics[width=\linewidth]{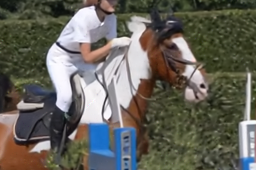}
					\end{subfigure}
					\begin{subfigure}[t]{.24\linewidth}
						\centering
							\includegraphics[width=\linewidth]{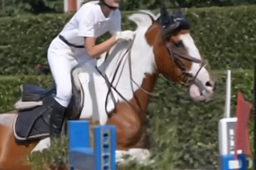}
					\end{subfigure}
					\begin{subfigure}[t]{.24\linewidth}
						\centering
							\includegraphics[width=\linewidth]{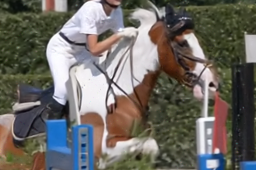}
					\end{subfigure}
				\subcaption{Ours - FastDVDnet ($\sqrt{\varepsilon}=30/255, \alpha=2.25$) PSNR = 29.55 dB}
			\end{subfigure}

			\caption{\label{fig:interpolation}Video interpolation of random missing pixels ($\rho=0.9, \sigma_{\mathbf{n}}=0$) after 200 iterations of video PnP-ADMM on the \texttt{horsejump-stick} sequence of DAVIS-2017-test-480p~\cite{Pont-Tuset_arXiv_2017}. Ours - DRUNet and FastDVDnet are using their respective optimal values of $(\varepsilon, \alpha)$. FastDVDnet produces significantly better, temporally more stable results.}
	\end{figure*}

	\noindent\textbf{Results.} Results of our method for interpolation of missing pixels for $\rho \in \left\{0.5, 0.9\right\}$ and $\sigma_{\mathbf{n}} \in \left\{0, 2.55, 7.65\right\}/255$ are shown in Table \ref{table:interpolation_davis}. We do not compare the performance of our method to DPIR here, as this problem is not part of the original publication (and thus there is no data fidelity subproblem nor optimal parameter values for their PnP-HQS algorithm). 
	\begin{table*}
		\caption{\label{table:interpolation_davis} Video interpolation of random missing pixels: PSNR/SSIM after 200 iterations of PnP-ADMM on DAVIS-2017-test-480p~\cite{Pont-Tuset_arXiv_2017} (each video limited to its 30 first frames)}
		\centering
		{\small
        % \begin{tabular}{ccccc}
        % \toprule
        %                             &                                & masked     & Ours - DRUNet                         & Ours - FastDVDnet                     \\
        % \midrule
        % \multirow{3}{*}{$\rho=0.5$} & $\sigma_{\mathbf{n}}=0$        & 9.44/0.142 & 44.73/\textbf{0.992} & \textbf{45.00}/0.991 \\
        %                             & $\sigma_{\mathbf{n}}=2.55/255$ & 9.43/0.138 & 39.89/0.970                           & \textbf{41.83/0.981} \\
        %                             & $\sigma_{\mathbf{n}}=7.65/255$ & 9.40/0.117 & 37.37/0.954                           & \textbf{38.08/0.960} \\
        % \midrule
        % \multirow{3}{*}{$\rho=0.9$} & $\sigma_{\mathbf{n}}=0$        & 6.88/0.047 & 26.14/0.820                           & \textbf{32.40/0.914} \\
        %                             & $\sigma_{\mathbf{n}}=2.55/255$ & 6.88/0.044 & 26.28/0.821                           & \textbf{32.22/0.910} \\
        %                             & $\sigma_{\mathbf{n}}=7.65/255$ & 6.86/0.032 & 26.70/0.813                           & \textbf{32.21/0.886}\\
        % \bottomrule
        % \end{tabular}
		
		\begin{tabular}{c| *{2}{c} c | *{2}{c} c}
			\toprule
             & \multicolumn{3}{c|}{$\rho=0.5$} & \multicolumn{3}{c}{$\rho=0.9$} \\
			%\cline{2-4} \cline{5-7}
			Method     & $\sigma_{\mathbf{n}}=0$     & $\sigma_{\mathbf{n}}=2.55/255$ & $\sigma_{\mathbf{n}}=7.65/255$ & $\sigma_{\mathbf{n}}=0$ & $\sigma_{\mathbf{n}}=2.55/255$     & $\sigma_{\mathbf{n}}=7.65/255$ \\
			\midrule
			masked           & 9.44/0.142 & 9.43/0.138 & 9.40/0.117 & 6.88/0.047 & 6.88/0.044 & 6.86/0.032\\
			Ours - DRUNet     & 44.73/\textbf{0.992} & 39.89/0.970 & 37.37/0.954 & 26.14/0.820 & 26.28/0.821 & 26.70/0.813\\
			Ours - FastDVDnet & \textbf{45.00}/0.991 & \textbf{41.83/0.981} & \textbf{38.08/0.960} & \textbf{32.40/0.914} & \textbf{32.22/0.910} & \textbf{32.21/0.886}\\
			\bottomrule
		\end{tabular}
		}
	\end{table*}
    In this experiment, since the amount of missing pixels and their locations change from one frame to another, the DRUNet-based scheme (equivalent to operating operating on each frame separately) clearly lags behind. The FastDVDnet-based scheme takes advantage of its neighbor-using denoising step to provide a much more reliable interpolation of the video. A visual example is also proposed in Figure \ref{fig:interpolation}. 
	\\
	
%	\subsection{Concluding remarks on parameter settings.}
%	\commentBy{AA}{L'ancienne conclusion gretsi trouve mieux sa place ici je trouve. Il s'agit de détails que l'on ne veut pas mettre tellement en avant}
%	According to our experiments, the more difficult the problem studied, the greater the number of iterations required. The convergence speed of the method and the final performance also depend on the content of the unknown and the parameters
%	$\varepsilon$ and $\alpha$, and this dependence varies with the denoiser used. Defining practices to ensure the stability of results is still an open question; for example, strategies for evolving $\varepsilon$ and $\alpha$ across iterations can be considered.

\newpage	
	\section{Conclusion}\label{sec:conclusion}
	In this work, we explored two strategies for restoring degraded videos using Plug \& Play algorithms:
	using \emph{(i)} a state-of-the-art image denoiser (DRUNet) as a regularizer, and
	\emph{(ii)} a lightweight, competitive video denoiser (FastDVDnet) as a regularizer.
	Whereas the first approach is equivalent to applying a PnP algorithm separately on each frame, the second approach is explored here for the first time.
	Our experiments show that the second approach outperforms the first one in the vast majority of problems tested (denoising, deblurring, super-resolution, missing pixels), both in terms of PSNR and temporal consistency. This performance difference is quite remarkable, as FastDVDnet has 13$\times$ less parameters than DRUNet. This suggests the possibility of even better performance were we to use more complex and evolved video denoisers like~\cite{Sun2021}. Promising results for video interpolation of random missing pixels paves the way for other video restoration problems where the degradation operator cannot be separated frame by frame, such as frame interpolation or spatio-temporal deconvolution. In future work, we also intend to provide convergence guarantees for video restoration PnP schemes under very mild conditions on the data-fitting term. A promising way to do so would be to generalize the work of~\cite{hurault_gradient_2022, Hurault2022b} to video denoisers.
	
	{\small
	\bibliographystyle{ieee_fullname}
	\bibliography{refs}
	}

\end{document}

\pagestyle{plain}
\pagenumbering{roman}

	%%%%%%%%% TITLE
	\title{Video Restoration with a Deep Plug-and-Play Prior: Supplementary material}

	\maketitle
	%\thispagestyle{empty}

% 	\begin{figure}
% 		\centering
% 			\includegraphics[width=\columnwidth]{images/cascaded_unets.png}
% 			\caption{FastDVDnet architecture~\cite{tassano_fastdvdnet_2020}\label{fig:fastdvdnet}}
% 			%		\begin{subfigure}[t]{\linewidth}
% 				%			\centering
% 					%				\includegraphics[width=\linewidth]{images/unet_FastDVDnet.png}
% 					%			
% 				%		\end{subfigure}
% 			%		\legende{Architecture d'un bloc de FastDVDnet}
% 	\end{figure}
% 	%
% 	\begin{figure}
% 		\centering
% 			\includegraphics[width=\columnwidth]{images/drunet.png}
% 			\caption{DRUNet architecture~\cite{zhang_plug-and-play_2021}\label{fig:Drunet}}
% 	\end{figure}

	\subsection*{Optimal setting of PnP parameters} \label{deblurring_params}
	
	For a given video restoration problem, our formulation of PnP-ADMM has several parameters that can have an impact on final performance:
	\begin{itemize}
		\item $\mathcal{D}_{\varepsilon}$, the Gaussian denoiser used as the proximal operator of the prior term.
		\item $\varepsilon$, which represents the strength of said denoiser. When the denoiser is designed to handle varying levels of noise in a non-blind fashion, (as is the case of FastDVDnet and DRUNet), $\varepsilon$ can be set in the form of a noise map that is fed as an additional input to the denoiser.
		\item $\alpha$, which represents the tradeoff between the data fidelity and prior terms of Equation (3).
		\item $K$, the total number of ADMM iterations. It needs to be large enough to allow convergence of the algorithm, but not too large in cases where convergence is not guaranteed.
	\end{itemize}
	For a given problem and a given denoiser (i.e. $\mathcal{A}$, $\sigma_{\mathbf{n}}$ and $\mathcal{D}$ are fixed), to find optimal values for said parameters, we perform a 2D grid search on $\varepsilon$ and $\alpha$ for an empirically found sufficient number of iterations $K$. Since we have no guarantees that these values are optimal for every possible $\mathbf{x}$, we perform that grid search on a subset of the data we wish to evaluate our method on (in our case, $256 \times 256$ center crops of the first 30 frames of the \texttt{aerobatics}, \texttt{girl-dog}, \texttt{horsejump-stick} and \texttt{subway} of DAVIS-2017-test-480p~\cite{Pont-Tuset_arXiv_2017}). We retain the $(\varepsilon, \alpha, K)$ tuple that yields the higher PSNR on that subset for a given denoiser. 
	\begin{figure}
		\centering
			% INCLUDE FIGURE
			\includegraphics[width=\columnwidth]{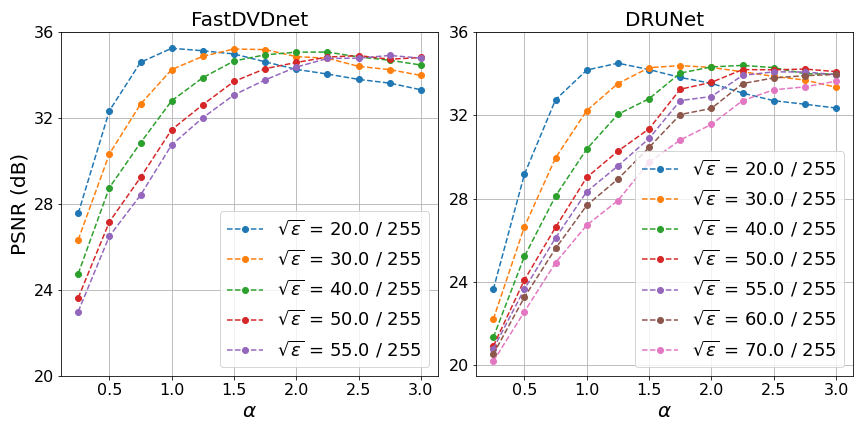}
			\caption{\label{fig:db_gridsearch}Grid search of optimal $(\varepsilon, \alpha)$ for video deblurring ($\sigma_{\mathbf{n}} = 2.55/255$)}
		
	\end{figure}
	An example of a grid search of $(\varepsilon, \alpha)$ for a given video deblurring problem is illustrated in Figure \ref{fig:db_gridsearch}. The parameter values obtained for our all video restoration problems studied in this work are given in Tables \ref{table:params_deblur}, \ref{table:params_sr}, \ref{table:params_interpolation} and \ref{table:params_demosaicking}.
%	\begin{itemize}
%		\item $(\sqrt{\varepsilon}, \alpha, K) = (20/255, 1, 20)$ for FastDVDnet at $\sigma_{\mathbf{n}}=2.55/255$,
%		\item $(\sqrt{\varepsilon}, \alpha, K) = (20/255, 0.5, 20)$ for FastDVDnet at $\sigma_{\mathbf{n}}=7.65/255$,
%		\item $(\sqrt{\varepsilon}, \alpha, K) = (20/255, 1.25, 10)$ for DRUNet at $\sigma_{\mathbf{n}}=2.55/255$,
%		\item $(\sqrt{\varepsilon}, \alpha, K) = (50/255, 1.25, 10)$ for DRUNet at $\sigma_{\mathbf{n}}=7.65/255$.
%	\end{itemize}
	\begin{table}
		\caption{\label{table:params_deblur}Video deblurring: selected values of $(\varepsilon, \alpha, K)$ PnP-ADMM parameters for $\sigma_{\mathbf{n}} \in \left\{ 2.55, 7.65 \right\} / 255$}
		\centering
		{\small
			\begin{tabular}{c c | c  c}
				\toprule
				                 
				\multicolumn{2}{c|}{} & $\sigma_{\mathbf{n}}=2.55$  & $\sigma_{\mathbf{n}}=7.65$ \\
				\multicolumn{2}{c|}{Denoiser} & $/255$  & $/255$ \\
				\midrule
				\multirow{3}{*}{DRUNet} & $\sqrt{\varepsilon} \times 255$ & 20 & 50\\
				 & $\alpha$ & 1.25 & 1.25 \\
				 & $K$ & 10 & 10 \\
				 \midrule
				\multirow{3}{*}{FastDVDnet} & $\sqrt{\varepsilon} \times 255$ & 20 & 20\\
				 & $\alpha$ & 1 & 0.5 \\
				 & $K$ & 20 & 20 \\
				\bottomrule
		\end{tabular}}
	\end{table}

% 	\subsection*{Super-resolution: optimal parameter setting} \label{sr_params}
	
% 	We use the methodology described in Section \ref{deblurring_params}, along with the same data subset, to find optimal parameters for 3 SR problems: $\times2$ super-resolution with a $25\times25$ Gaussian blur kernel with standard deviation of 1.6, $\times4$ super-resolution with the same kernel, and $\times4$ super-resolution with a $25\times25$ Gaussian blur kernel with standard deviation of 3.2 (which produces a less aliased LR image). Additionally, we evaluate each of these problems  at 2 noise levels: $\sigma_{\mathbf{n}}=0$ and $\sigma_{\mathbf{n}}=2.55/255$. The resulting parameter values after grid search are shown in Table \ref{table:params_sr}.

	\begin{table*}
		\caption{\label{table:params_sr}Video super-resolution: selected values of $(\varepsilon, \alpha, K)$ PnP-ADMM parameters}
		\centering
		{\small
			\begin{tabular}{c c |*{2}{c c|} c c}
				\toprule
				
				\multicolumn{2}{c|}{s.f. / kernel} & \multicolumn{2}{c|}{$\times2$/ Gauss. ($\sigma=1.6$)} & \multicolumn{2}{c|}{$\times4$ / Gauss. ($\sigma=1.6$)} & \multicolumn{2}{c}{$\times4$ / Gauss. ($\sigma=3.2$)} \\
				\multicolumn{2}{c|}{Denoiser} & $\sigma_{\mathbf{n}}=0$     & $\sigma_{\mathbf{n}}=2.55/255$ & $\sigma_{\mathbf{n}}=0$ & $\sigma_{\mathbf{n}}=2.55/255$ & $\sigma_{\mathbf{n}}=0$     & $\sigma_{\mathbf{n}}=2.55/255$ \\
				\midrule
				\multirow{3}{*}{DRUNet} & $\sqrt{\varepsilon} \times 255$ & 40 & 55 & 60 & 60 & 60 & 70\\
				& $\alpha$ & 0.075 & 1.25  & 1 & 1.25 & 0.025 & 0.25 \\
				& $K$ & 10 & 10 & 10 & 10 & 10 & 10 \\
				\midrule
				\multirow{3}{*}{FastDVDnet} & $\sqrt{\varepsilon} \times 255$ & 20 & 30 & 55 & 55 & 50 & 55\\
				& $\alpha$ & 0.025 & 0.5 & 0.75 & 0.75 & 0.025 & 0.25 \\
				& $K$ & 20 & 20 & 20 & 20 & 20 & 20 \\
				\bottomrule
		\end{tabular}}
	\end{table*}

% 	\subsection*{Interpolation: optimal parameter setting} \label{interpolation_params}

%     We apply the methodology of Sections \ref{deblurring_params}, \ref{sr_params} and \ref{demosaicking_params} to find optimal $(\varepsilon, \alpha)$ for a sufficiently large number of iterations $K=200$. We evaluate our interpolation problem for $\rho \in \left\{0.5, 0.9\right\}$ and $\sigma \in \left\{0, 2.55, 7.65\right\}/255$. The parameter values are shown in Table~\ref{table:params_interpolation}.
    %
	\begin{table*}
		\caption{\label{table:params_interpolation}Video interpolation of random missing pixels: selected values of $(\varepsilon, \alpha, K)$ PnP-ADMM parameters}
		\centering
		{\small
			\begin{tabular}{c c | *{2}{c} c | *{2}{c} c}
				\toprule
				
				\multicolumn{2}{c|}{} & \multicolumn{3}{c|}{$\rho=0.5$} & \multicolumn{3}{c}{$\rho=0.9$}  \\
				\multicolumn{2}{c|}{Denoiser} & $\sigma_{\mathbf{n}}=0$     & $\sigma_{\mathbf{n}}=2.55/255$ & $\sigma_{\mathbf{n}}=7.65/255$ & $\sigma_{\mathbf{n}}=0$ & $\sigma_{\mathbf{n}}=2.55/255$     & $\sigma_{\mathbf{n}}=7.65/255$ \\
				\midrule
				\multirow{3}{*}{DRUNet} & $\sqrt{\varepsilon} \times 255$ & 30 & 30 & 30 & 50 & 50 & 50 \\
				& $\alpha$ & 0.5 & 3.0 & 2.5 & 2.75 & 2.5 & 2.75 \\
				& $K$ & 200 & 200 & 200 & 200 & 200 & 200 \\
				\midrule
				\multirow{3}{*}{FastDVDnet} & $\sqrt{\varepsilon} \times 255$ & 20 & 20 & 20 & 30 & 30 & 40 \\
				& $\alpha$ & 1.75 & 3.0 & 1.5 & 2.25 & 2.75 & 1.5 \\
				& $K$ & 200 & 200 & 200 & 200 & 200 & 200 \\
				\bottomrule
		\end{tabular}}
	\end{table*}

%DRUNet 	 prob = 0.5, sigma = 0.0 	 best epsilon = 30.0   best alpha = 0.5 	 PSNR = 43.25974655151367
%DRUNet 	 prob = 0.5, sigma = 2.55 	 best epsilon = 30.0   best alpha = 3.0 	 PSNR = 39.379119873046875
%DRUNet 	 prob = 0.5, sigma = 7.65 	 best epsilon = 30.0   best alpha = 2.5 	 PSNR = 35.91905212402344
%DRUNet 	 prob = 0.9, sigma = 0.0 	 best epsilon = 50.0   best alpha = 2.75 	 PSNR = 27.852909088134766
%DRUNet 	 prob = 0.9, sigma = 2.55 	 best epsilon = 50.0   best alpha = 2.5 	 PSNR = 27.85205078125
%DRUNet 	 prob = 0.9, sigma = 7.65 	 best epsilon = 50.0   best alpha = 2.75 	 PSNR = 27.655309677124023
%FastDVDnet 	 prob = 0.5, sigma = 0.0 	 best epsilon = 20.0   best alpha = 1.75 	 PSNR = 44.06608581542969
%FastDVDnet 	 prob = 0.5, sigma = 2.55 	 best epsilon = 20.0   best alpha = 3.0 	 PSNR = 40.5982666015625
%FastDVDnet 	 prob = 0.5, sigma = 7.65 	 best epsilon = 20.0   best alpha = 1.5 	 PSNR = 36.326416015625
%FastDVDnet 	 prob = 0.9, sigma = 0.0 	 best epsilon = 30.0   best alpha = 2.25 	 PSNR = 31.328664779663086
%FastDVDnet 	 prob = 0.9, sigma = 2.55 	 best epsilon = 30.0   best alpha = 2.75 	 PSNR = 31.137882232666016
%FastDVDnet 	 prob = 0.9, sigma = 7.65 	 best epsilon = 40.0   best alpha = 1.5 	 PSNR = 30.17365264892578

	\subsection*{Bayer to RGB video demosaicking}\label{seq:demosaicking}

    In addition to deblurring, super-resolution and interpolation of random missing pixels, we evaluate a very recurring topic of digital photography: demosaicking~\cite{gunturk2005demosaicking}. Indeed, the vast majority of CMOS image sensors currently used in smartphones, cameras and various computer vision systems are monochrome sensors stacked with an array of per-pixel color filters (called Color Filter Array / CFA). That array is typically the repetition of a $2\times2$ pattern, where two green color filters are put diagonally and the other two filters are red and blue (called "Bayer" pattern). This means that the sensor output is undersampled in terms of colors at each pixel location, and an interpolation algorithm must be used to recover red, green and blue values at each pixel. Videos are of course not spared from that problem since they also come from CMOS image sensors. The degradation model for a synthetic video demosaicking inverse problem can be written as
	\begin{equation}\label{eq:forward_demosaicking}
		\mathbf{y} = \mathbf{M} \odot \mathbf{x} + \mathbf{n}
	\end{equation}
	where $\mathbf{M}$ is a binary mask corresponding to the Bayer CFA pattern and $\odot$ denotes pixel-wise multiplication. Notice that this is actually the same degradation model as the interpolation problem of Section 4.3 (Equation (15)) with a a different, fixed mask.	The proximal operator of the data term is thus identical to the expressions of Equations (16) and (17).
    \newline
	
	\noindent\textbf{Parameter setting of the demosaicking problem.} For this specific problem only, we experiment with a variant of our method of video PnP-ADMM: an annealed version, where the strength of the denoiser $\varepsilon$ changes accross iterations. We are not the first to consider that kind of strategy, as variants of ADMM where the penalty parameter is updated at each iteration are already mentioned in~\cite{boyd_distributed_2011}, and similar annealing strategies were used both in PnP-ADMM~\cite{brifman_turning_2016, chan_plug-and-play_2016} and other schemes such as PnP-SGD~\cite{laumont_maximum--posteriori_2021} and PnP-HQS~\cite{zhang_plug-and-play_2021}.
	
\begin{figure}
		\centering
		\includegraphics[width=\columnwidth]{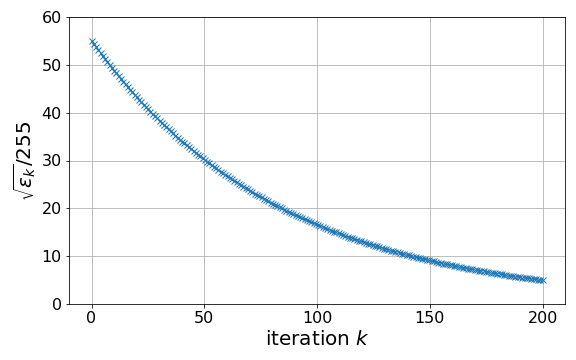}
		\caption{\label{fig:eps_annealing}Example of annealing of $\varepsilon$ with $\sqrt{\varepsilon_{k}} \in \left[55, 5\right] / 255$ and $K=200$}
	\end{figure}

	We have seen in other restoration problems that the best value of $\alpha$ for a given denoiser and a given problem typically depends on the selected value of $\varepsilon$. This could be a problem for our annealing strategy, since we want $\varepsilon$ to change across iterations (we might have to change $\alpha$ across iterations as well, making our 2D grid search strategy impossible in that case). Luckily, as shown in Equation (17) of Section 4.3, in both masking problems, when there is no noise (or little noise), the proximal operator of the data term does not depend on $\alpha$, allowing us to consider our strategy of annealing of $\varepsilon$ while keeping the same $\alpha$ across iterations. Similarly to what is done in~\cite{zhang_plug-and-play_2021}, we uniformly sample $\sqrt{\varepsilon}$ from a large noise level $\sqrt{\varepsilon}_{0}$ to a small one $\sqrt{\varepsilon_{K-1}}$ in log space. We set $\sqrt{\varepsilon_{K-1}} = 5/255$ for FastDVDnet since it was trained for noise levels in $[55, 5]/255$, and appropriate denoising performance is not guaranteed at noise levels outside of this range (as shown in Table 1 of the main paper). Given that DRUNet was trained for noise levels in $[50, 0]/255$ and that it generalizes better to unseen noise levels, we set $\sqrt{\varepsilon_{K-1}} = 1/255$ when using annealing with that network. An example of annealing of $\varepsilon$ for $K=200$ iterations (an empirically found sufficient number of iterations for PnP-ADMM applied to video demosaicking) is shown in Figure \ref{fig:eps_annealing}.

    As done in other problems, we thus perform our grid search of optimal values of $\varepsilon$ and $\alpha$ for FastDVDnet and DRUNet for Bayer RGGB to RGB demosaicking, at three noise levels $\sigma_{\mathbf{n}}=0$, $\sigma_{\mathbf{n}}=2.55/255$ and $\sigma_{\mathbf{n}}=7.65/255$. We perform that grid search both with and without annealing of $\varepsilon$ to later compare the performance of both PnP-ADMM versions. The resulting parameter values are shown in Table \ref{table:params_demosaicking}.
	\begin{table}
		\caption{\label{table:params_demosaicking}Video demosaicking: selected values of $(\varepsilon, \alpha, K)$ PnP-ADMM parameters for $\sigma_{\mathbf{n}} \in \left\{ 0, 2.55, 7.65 \right\} / 255$}
		\centering
		{\small
			\begin{tabular}{c c | *{2}{c} c}
				\toprule
				
				\multicolumn{2}{c|}{}  &$\sigma=0$ & $\sigma_{\mathbf{n}}=2.55$  & $\sigma_{\mathbf{n}}=7.65$ \\
				\multicolumn{2}{c|}{Denoiser}&  & $/255$  & $/255$ \\ \midrule
				\multirow{3}{*}{DRUNet} & $\sqrt{\varepsilon} \times 255$ & 60 & 60 & 60\\
				& $\alpha$ & 0.25 & 2.5 & 3 \\
				& $K$ & 200 & 200 & 200 \\
				\midrule
				\multirow{3}{*}{FastDVDnet} & $\sqrt{\varepsilon} \times 255$ & 20 & 20 & 20\\
				& $\alpha$ & 0.25 & 3 & 1.5 \\
				& $K$ & 200 & 200 & 200 \\
				\midrule
				& $\sqrt{\varepsilon_{0}} \times 255$ & 70 & 70 & 70\\
				DRUNet & $\alpha$ & 0.25 & 0.75 & 0.5 \\
				(annealing) & $K$ & 200 & 200 & 200 \\
				\midrule
				& $\sqrt{\varepsilon_{0}} \times 255$ & 30 & 55 & 50\\
				FastDVDnet & $\alpha$ & 0.25 & 1 & 0.5 \\
				(annealing)& $K$ & 200 & 200 & 200 \\
				\bottomrule
		\end{tabular}}
	\end{table}

\begin{figure}
		\centering
		\includegraphics[width=\columnwidth]{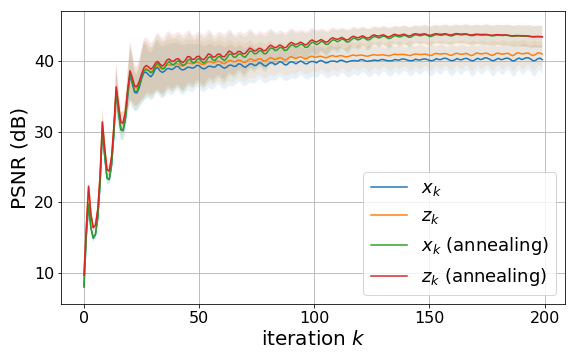}
		\caption{\label{fig:psnr_iter_annealing}Demosaicking ($\sigma_{\mathbf{n}}=0$): PSNR of split variables across iterations $\mathbf{x}_{k}$, $\mathbf{z}_{k}$ with and without annealing (FastDVDnet, $\sqrt{\varepsilon_{0}} = \sqrt{\varepsilon} = 20 /255$). Results are computed and averaged on $256 \times 256$ crops of the first 30 frames of sequences \texttt{aerobatics}, \texttt{girl-dog}, \texttt{horsejump-stick} and \texttt{subway} of DAVIS-2017-test-480p~\cite{Pont-Tuset_arXiv_2017}}
	\end{figure}

	Besides from potentially increasing performance, another added benefit of decreasing $\varepsilon$ across iterations is that it forces the distance between $\mathbf{x}$ and $\mathbf{z}$ to decrease when trying to minimize the augmented Lagrangian (Equation (4)), yielding additional stability of results in cases where the convergence of ADMM is not guaranteed. This is illustrated in Figure~\ref{fig:psnr_iter_annealing}.
 	\newline
	
	\noindent\textbf{Results.} As done in Sections 4.1 and 4.2, we include DPIR~\cite{zhang_plug-and-play_2021} to our performance comparison, with the parameter values recommended by the authors for demosaicking: $(\sigma_{1},\sigma_{K}, \lambda, K) = (49, \max(0.6, \sigma_{\mathbf{n}}*255), 0.23, 40)$ using the notations of~\cite{zhang_plug-and-play_2021} (we disable here again the periodical geometric self-ensemble). It is interesting to note that for our method, we initialize with the mosaicked observation: $\mathbf{z}_{0}=\mathbf{y}$. In contrast, DPIR initializes to an already crudely demosaicked version of the observation (the authors of~\cite{zhang_plug-and-play_2021} use matlab's \texttt{demosaic} function). We do not do so, as we only found marginal improvements to our PnP-ADMM results both in term of final performance and convergence speed. A drawback is that our method typically requires more iterations (100 to 200) than DPIR with its matlab initialization (40). Conversely, we found in our experiments that DPIR would also require more iterations with a mosaicked initialization, but we leave that out of our performance comparison, as no number of iterations $K$ was provided by its authors for such a use case. Quantitative results of our PnP-ADMM methods on the 2017 test test of the DAVIS dataset~\cite{Pont-Tuset_arXiv_2017} are available in Table~\ref{table:demosaicking_davis}.
	\begin{table}
		\caption{\label{table:demosaicking_davis}Video demosaicking: PSNR/SSIM on DAVIS-2017-test-480p~\cite{Pont-Tuset_arXiv_2017} for $\sigma_{\mathbf{n}} \in \left\{ 0, 2.55, 7.65 \right\} / 255$ (each video limited to its 30 first frames)}
		\centering
		{\small
			\begin{tabular}{c| *{2}{c} c}
				\toprule
	                   & $\sigma_{\mathbf{n}}=0$  & $\sigma_{\mathbf{n}}=2.55$ & $\sigma_{\mathbf{n}}=7.65$\\
				Method             &  & $/255$ & $/255$\\
				\midrule
				\multirow{2}{*}{mosaicked}          & \multirow{2}{*}{8.21/0.100} & \multirow{2}{*}{8.21/0.096} & \multirow{2}{*}{8.18/0.079}\\
				& & & \\
%					matlab init        & 22.75/0.5994 & 12.66/0.533 &\\
				\multirow{2}{*}{Ours - DRUNet}       & \multirow{2}{*}{33.23/0.921} & \multirow{2}{*}{32.20/0.879} & \multirow{2}{*}{31.22/0.847}\\
				& & & \\
				\multirow{2}{*}{Ours - FastDVDnet}   & \multirow{2}{*}{39.78/0.971} & \multirow{2}{*}{38.78/0.964} & \multirow{2}{*}{36.68/\textbf{0.952}}\\
				& & & \\
				Ours - DRUNet & \multirow{2}{*}{45.15/0.992} & \multirow{2}{*}{40.55/0.973} & \multirow{2}{*}{\textbf{36.75}/0.951}\\
				(annealing) & & &\\
				Ours - FastDVDnet & \multirow{2}{*}{43.92/0.990} & \multirow{2}{*}{\textbf{41.04/0.981}} & \multirow{2}{*}{36.40/0.951}\\
				(annealing) & & & \\
				\multirow{2}{*}{DPIR~\cite{zhang_plug-and-play_2021}} & \multirow{2}{*}{\textbf{45.72/0.993}} & \multirow{2}{*}{40.05/0.972} & \multirow{2}{*}{34.48/0.905}\\
				& & & \\
				\bottomrule
		\end{tabular}}
	\end{table}
    These results show the clear benefit of using an annealed version of PnP-ADMM for demosaicking. Similarly to the other restoration problems, they also show that FastDVDnet performs better than DRUNet in most scenarios. It is worh noting that DPIR performs better than our method (with annealing and either denoiser) in the noiseless case, but worse than our method (with annealing and either denoiser) at $\sigma_{\mathbf{n}}=7.65/255$. %Similarly to the previous restoration problem, PnP-ADMM with FastDVDnet produces the best results in most cases.

    \subsection*{Runtimes}
    
    Plug-and-play methods are typically not real-time methods, because they imply multiple iterations of an alternate optimization algorithm, and each iteration itself includes a forward pass through a denoising algorithm. Our method is no exception. That said, our method is typically faster than DPIR when performing video restoration with similar initialization, since the number of iterations required by DPIR for a single frame must be multiplied by the number of frames of the video. In contrast, one iteration of our method processes the whole video at once. Actual runtimes of course depend on the required number of iterations, which can in turn depend on the scene content, the difficulty of the problem, the type of initialization and the choice of parameters. Here are the runtimes with the parameters described in the article (8 CPU AMD, 1 NVIDIA A100, 64GB RAM), DAVIS-2017-test-480p~\cite{Pont-Tuset_arXiv_2017}):
    \begin{itemize}
    	\small
    	\item \textbf{deblurring}: Ours - FastDVDnet 1.2s/frame (20 it.), Ours - DRUNet 1.4s/frame (10 it.), DPIR 2.7s/frame (8 it.)
    	\item \textbf{sr}: Ours - FastDVDnet 1.2s/frame (20 it.), Ours - DRUNet 1.4s/frame (10 it.), DPIR 8.1s/frame (24 it.)
    	\item \textbf{interpolation}: Ours - FastDVDnet 12s/frame (200 it.), Ours - DRUNet 29s/frame (200 it.)
    	\item \textbf{demosaicking}: Ours - FastDVDnet 12s/frame (200 it.), Ours - DRUNet 29s/frame (200 it.) , DPIR 14s/frame (40 it.)
    \end{itemize}

	{\small
	\bibliographystyle{ieee_fullname}
	\bibliography{refs}
	}